\documentclass[a4paper,11pt]{article}
\usepackage{jheppub} 
\usepackage{amsthm}
\usepackage{amssymb}  
\usepackage{caption}
\usepackage{feynmf}
\usepackage{stackengine} 
\usepackage{breqn} 
\usepackage{float}
\usepackage[normalem]{ulem}
\usepackage{cancel}
\newtheoremstyle{mystyle}
{3pt}
{3pt}
{}
{}
{}
{.}
{.5em}
{}

\ifpdf
\fi

\makeatletter
\def\endfmffile{%
	\fmfcmd{\p@rcent\space the end.^^J%
		end.^^J%
		endinput;}%
	\if@fmfio
	\immediate\closeout\@outfmf
	\fi
	\IfFileExists{\thefmffile.mp}{\immediate\write18{mpost \thefmffile}}{}
	\let\thefmffile\relax
}
\makeatother

\def\a{\alpha}
\def\b{\beta}
\def\d{\delta}

\def\ve{\varepsilon}

\def\g{\gamma}
\def\h{\eta}
\def\j{\psi}

\def\l{\lambda}
\def\m{\mu}
\def\n{\nu}
\def\o{\omega}
\def\p{\pi}

\def\r{\rho}
\def\s{\sigma}
\def\t{\tau}

\def\x{\xi}

\def\D{\Delta}

\def\G{\Gamma}

\def\O{\Omega}
\def\P{\Pi}

\def\S{\Sigma}

\def\pa{\partial}

\newcommand{\ov}{\overline}
\DeclareMathOperator{\Tr}{Tr}
\newcommand{\wt}{\widetilde}

\def\Sl#1{\rlap{\hbox{$\mskip 3 mu /$}}#1}
\newcommand{\pari}{\stackrel{{P}}\longrightarrow}
\newcommand{\wh}{\widehat}
\newcommand{\stkout}[1]{\ifmmode\text{\sout{\ensuremath{#1}}}\else\sout{#1}\fi}
\newtheorem{mydef}{Definition}

\title{\boldmath 
The BPHZL renormalization of a planar quantum electrodynamics up to 2-loops and beyond}

\author[a]{D.O.R. Azevedo,}
\author[a,d]{O.M. Del Cima,}
\author[b]{L.S. Lima}
\author[a,c]{and E.S. Miranda}
\affiliation[a]{Departamento de F\'{i}sica (DPF) \\ Universidade Federal de Vi\c{c}osa (UFV)\\
	Avenida Peter Henry Rolfs s/n - 36570-900 - Vi\c cosa - MG - Brazil.}
	
\affiliation[b]{Departamento de F\'{i}sica (CODAFIS) \\ Instituto Federal de Educa\c c\~ao, Ci\^encia e Tecnologia de Minas Gerais (IFMG) - Campus Ouro Preto\\
	Rua Pandi\'{a} Cal\'{o}geras, 898 - Bauxita - 35400-000 - Ouro Preto - MG - Brazil}
	
\affiliation[c]{Secretaria de Estado de Educa\c c\~ao de Minas Gerais \\
Superintend\^encia Regional de Ensino de Ub\'a \\
Av. Raul Soares, 30 - Centro - 36500-000 - Ub\'a - MG - Brazil}

\affiliation[d]{Ibitipoca Institute of Physics (IbitiPhys),\\
36140-000 - Concei\c c\~ao do Ibitipoca - MG - Brazil.}

\emailAdd{daniel.azevedo@ufv.br, oswaldo.delcima@ufv.br, lazaro.lima@ifmg.edu.br, emerson.silva.miranda@educacao.mg.gov.br}

\abstract{The renormalization of a parity-even massless $U(1)\times U(1)$ quantum electrodynamics in three space-time dimensions (QED$_3$) is studied by adopting the Bogoliubov-Parasiuk-Hepp-Zimmermann-Lowenstein (BPHZL) renormalization method. The presence of two massless fermions requires the Lowenstein-Zimmermann (LZ) subtraction scheme to renormalize the infrared divergences induced by the ultraviolet subtractions at 1- and 2-loops, moreover due to the model superrenormalizability the ultraviolet divergences are bounded up to 2-loops. Finally, through the BPHZL renormalization method together with the LZ subtraction scheme the ultraviolet and infrared finiteness of the model is proved up to 2-loops and beyond. }

\begin{document}
	\maketitle
	\flushbottom
	
	\section{Introduction}
	\label{sec:intro}

Quantum electrodynamics in three space-time dimensions (QED$_3$), often referred to as planar Quantum Electrodynamics, has attracted significant attention since the seminal works of Schonfeld, Jackiw, Templeton, and Deser \cite{Schonfeld,Deser-Jackiw-Templeton,Jackiw-Templeton,Deser-Jackiw-Templeton-1982}. It has been considered as a framework for describing various quasi-planar condensed matter phenomena, including the Quantum Hall Effect \cite{Laughlin,Schakel,Kaplan-Sen}, high-$T_C$ superconductivity \cite{Herbut,Vafek,Christiansen:2002ma}, topological insulators \cite{Hasan-Kane,Kane_2011}, and graphene \cite{Gusynin,Jackiw-Pi,Katsnelson-Novoselov}.

In particular, the parity-preserving massless $U(1)\times U(1)$ Maxwell-Chern-Simons QED$_3$ -- referred to as the parity-preserving extended QED$_3$ throughout this work -- exhibits relevant features, such as the four-fold broken degeneracy of the Landau levels for electron-polaron and hole-polaron quasiparticles \cite{DeLima}, a phenomenon observed experimentally in pristine graphene \cite{zhang}. Furthermore, the model presents a Landau level with zero energy, which suggests the possibility of an anomalous Quantum Hall Effect \cite{DeLima}. This zero-energy Landau level has been observed experimentally in \cite{checkelski}. Moreover, the model is free from anomalies, and their ultraviolet and infrared finiteness emulate the scale invariance observed in graphene \cite{DelCima-Lima-Miranda}.

The ordinary massless $U(1)$ QED$_3$ is infrared and ultraviolet finite, parity and infrared anomaly free at all radiative order \cite{finiteness-qed}, however at 1-loop parity is explicitly broken in the course of Lowenstein-Zimmermann (LZ) infrared subtractions in the Bogoliubov-Parasiuk-Hepp-Zimmermann-Lowenstein (BPHZL) renormalization scheme \cite{bphzl}, consequently the 1-loop radiatively induced parity-odd Chern-Simons term to the vacuum-polarization
tensor is nothing but a counterterm owing to parity-violating LZ infrared subtractions in the BPHZL program. Meanwhile, a fundamental question has arisen if regardless of model the LZ subtraction scheme would necessarily violate parity in three dimensional space-time, {\it i.e.} if whether or not parity should be broken at any perturbative order throughout the infrared subtractions within the BPHZL renormalization scheme. The answer to this issue was clarified in \cite{parity}, where by analyzing the parity properties of the BPHZL subtraction operators for the model renormalized here -- the parity-even massless $U(1)\times U(1)$ Maxwell-Chern-Simons QED$_3$ -- it has been demonstrated that the parity is not broken at one and two loops. Consequently, if whether or not parity is broken, by the subtractions of infrared divergences induced by the ultraviolet ones,  depends on the model, namely the LZ infrared subtractions might induce or not parity violation in three space-time dimensions. Finally, to best of our knowledge this is new in quantum field theory for a discrete symmetry.
    
The proof on the BPHZL renormalization up to two loops of the parity-preserving massless $U(1)\times U(1)$ Maxwell-Chern-Simons QED$_3$, and as a byproduct its ultraviolet and infrared finiteness, is organized as follows. In Section 2 the action of the model is introduced and some useful gamma matrices relations are established. Moreover, in Section 2.1 the classical symmetries and Feynman rules are presented, in Section 2.2 the ultraviolet and infrared dimensions of all the fields and the power counting are established, and in Section 2.3 all Feynman diagrams at 1- and 2-loops are displayed. The BPHZL subtraction scheme, and 1-loop polarization tensor and self-energy are presented in Section 3. In Section 4, the expressions for the ultraviolet divergent graphs at 2-loops are written. The Zimmermann's forests are built up for the BPHZL renormalization at 2-loops in Section 4.1. The 2-loop BPHZL renormalization of diagrams $\g_{11_\pm}$ and $\g_{12_\pm}$, and of      
$\g_{8_\pm}$ and $\g_{9_\pm}$, are left to Section 4.2 and 4.3, respectively. The Appendix A presents some notations and useful relations, in the Appendix B the loop integrals (J$_r$-integrals) are displayed and a new general expression (B.7) for any number of momenta in the integrand numerator is determined, and in the Appendix C the $X'_l$ terms associated to the 2-loop diagrams $\g_{8_\pm}$ and $\g_{9_\pm}$ are explicitly computed.


\section{The parity-preserving massless $U(1)\times U(1)$ Maxwell-Chern-Simons QED$_3$}
\quad
In this section we present the main features of the parity-even massless $U(1)\times U(1)$ QED$_3$ proposed in \cite{DeLima}, where we have already incorporated a parity-even Lowenstein-Zimmermann mass term, which are required for renormalizing massless theories in BPHZL framework \cite{lz1,lz2}. A detailed discussion about this renormalizaion procedure will take place in the next sections. It is worth noting that the version of the action used here is the same as that presented in \cite{parity}, and the original one \cite{DeLima} is recovered when we take the limit $s \rightarrow 1$. The action reads
\begin{dmath}
	\S^{(s-1)} = \int{d^3 x} \bigg\{-\frac{1}{4}F^{\m\n}F_{\m\n} -\frac{1}{4}f^{\m\n}f_{\m\n}+ \m \ve^{\m\a\n}A_\m\pa_\a a_\n + 
	i {\ov\j_+} {\Sl D}\j_+ + i {\ov\j_-} {\Sl D}\j_- 
	\underbrace{-\,m(s-1){\ov\j_+}\j_+ +m(s-1){\ov\j_-}\j_-}_{ \textrm{\small Lowenstein-Zimmermann mass term}}+\, b \pa^\m A_\m+\frac{\a}{2}b^2+\ov{c}\square c+\pi \pa^\m a_\m+\frac{\b}{2}\pi^2+\ov{\x}\square \x \bigg\}~. \label{action} 
\end{dmath}

The covariant derivatives are defined as ${\Sl D}\j_\pm \!\equiv\!(\Sl\pa + ie\Sl{A} \pm ig\Sl{a})\j_\pm$, where $m$ and $\m$ are mass parameters with mass dimension equal to 1. Additionally, the coupling constants are $e$ (electric charge) and $g$ (pseudochiral charge), with mass dimension equal to 1/2. The field strengths
$F_{\m\n}=\pa_\mu A_\nu - \pa_\n A_\m$ and $f_{\m\n}=\pa_\mu a_\nu - \pa_\n a_\m$ are related to the electromagnetic field ($A_\m$) and the pseudochiral gauge field ($a_\m$), respectively. Furthermore, the Dirac spinors $\j_+$ and $\j_-$ are two kinds of fermions, associated to the two different sublattices of graphene, and the subscript $\pm$ is associated to the sign of the pseudospin \cite{DeLima,Binegar}. Moreover, the fields $c$ and $\x$
are two kinds of ghosts, whereas $\ov{c}$ and $\ov{\x}$ are two anti-ghosts. The fields $b$ and $\pi$ are Lautrup-Nakanishi fields \cite{lautrup-nakanishi}, which play a role of Lagrange multipliers, fixing the gauge. Finally, the Lowenstein-Zimmermann parameter $s$ take values from $0$ to $1$ and plays the same
role of the external momentum in the BPHZL renormalization procedure as a subtraction
variable. 

The representation used for the $\g$ matrices in three-dimensional space-time is $\g^\m=(\s_z,-i\s_x,i\s_y)$, and below are some of their properties that will be useful:
\begin{align}
	&\g^\m\g^\n=\h^{\m\n} {\mathbb I}+i\ve^{\m\n\a}\g_\a~,~~{\rm Tr}\{\g^\m \g^\n\}=2\h^{\m\n}~,~~{\rm Tr}\{\g^\m\g^\n\g^\a\}=2i\ve^{\m\n\a}~, \nonumber\\  
	&{\rm Tr}\{\g^{\m_1}\cdots \g^{\m_n}\}=
	\h^{\m_{n-1}\m_n}{\rm Tr}\{\g^{\m_1}\cdots \g^{\m_{n-2}}\}+i\ve^{\m_{n-1}\m_n\a}
	{\rm Tr}\{\g^{\m_1}\cdots \g^{\m_{n-2}}\g_\a\}~.\label{identities}
\end{align}


\subsection{Classical symmetries and Feynman rules}
\quad \,
The action $\S^{(s-1)}$ (\ref{action}) 
is invariant under the Becchi-Rouet-Stora (BRS) symmetry:
\begin{align}
&s\j_+=i(c + \x)\j_+~,~~s\ov{\j}_+=-i(c + \x)\ov{\j}_+~,& \nonumber \\
&s\j_-=i(c - \x)\j_-~,~~s\ov{\j}_-=-i(c - \x)\ov{\j}_-~,& \nonumber \\
&\displaystyle sA_\m=-\frac{1}{e}\pa_\m c~,~~s c=0~,~~ \displaystyle sa_\m=-\frac{1}{g}\pa_\m \x~,~~s\x=0~,& \nonumber \\
&\displaystyle s\ov{c}=\frac{b}{e}~,~~sb=0~,~~ \displaystyle s\ov{\x}=\frac{\pi}{g}~,~~s\pi=0~.&   \label{BRS}
\end{align}
For a detailed discussion of the BRS symmetry, see the refs. \cite{brs1,brs2,Piguet_Sorella}.

The action is also invariant under the parity symmetry:
\begin{align}
	\nonumber
	& \j_+ \pari \j_+^P=-i\g^1\j_-~,~~ \j_- \pari \j_-^P=-i\g^1\j_+~, \nonumber\\ 
	&\ov\j_+ \pari \ov\j_+^P=i\ov\j_-\g^1~,~~ \ov\j_- \pari \ov\j_-^P=i\ov\j_+\g^1~,\nonumber\\
	& A_\mu \pari A_\mu^P=(A_0,-A_1,A_2)~,~~ \phi \pari \phi^P=\phi~,~~\phi=\{b, c, \ov{c}\}~,\nonumber\\
	& a_\mu \pari a_\mu^P=(-a_0,a_1,-a_2)~,~~ \chi \pari \chi^P=-\chi~,~~\chi=\{\pi, \x, \ov{\x}\}~.\label{parity_transformation}
\end{align}
The propagators in momentum space are
\begin{subequations}
	\begin{align}\label{proaa}
		& \D^{\m\n}_{AA}(k) = -i\bigg\{\frac{1}{k^2-\mu^2}\left(\h^{\m\n}-\frac{k^\m k^\n}{k^2}\right)+\frac{\a}{k^2}\frac{k^\m k^\n}{k^2}\Bigg\}~,~~
	\end{align}  
	\begin{align}       
		&\D^{\m\n}_{aa}(k) = -i\Bigg\{\frac{1}{k^2-\mu^2}\left(\h^{\m\n}-\frac{k^\m k^\n}{k^2}\right)+\frac{\b}{k^2}\frac{k^\m k^\n}{k^2}\Bigg\}~,~~~
	\end{align}
	\begin{align}
		& \D_{Aa}^{\m\n}(k) = \frac{\mu}{k^2(k^2-\mu^2)}\ve^{\mu\a\n}k_\a~,~~\D_{Ab}^\m(k) = \D_{a \pi}^\m(k) = \frac{k^\m}{k^2}~,~~
	\end{align}
	\begin{align}
		& \D_{bb}(k) = \D_{\pi\pi}(k) = 0~,~~\D_{\ov{c}c}(k) = \D_{\ov{\x} \x}(k) = -\frac{i}{k^2}~, ~~~~~~~
	\end{align}
	\begin{align}        
		& \D_{++}(k) = i\frac{{\Sl k}-m(s-1)}{k^2-m^2(s-1)^2}~,~ \D_{--}(k) = i\frac{{\Sl k}+m(s-1)}{k^2-m^2(s-1)^2}~.  \label{propk}             
	\end{align}
\end{subequations}

From now on, all the calculations will be performed considering the Landau gauge,
where  $\a=\b=0$.

The diagramatic convensions for the propagators are:
\begin{fmffile}{feynmanrules}
	\begin{equation}
		\Delta_{AA}^{\mu \nu} \equiv
		\parbox{40pt}{
			\begin{fmfgraph*}(13,10)
				\fmfleft{i}
				\fmfright{o}
				\fmf{photon}{i,o}
		\end{fmfgraph*}}\quad,\quad
		\Delta_{aa}^{\mu \nu} \equiv
		\parbox{40pt}{
			\begin{fmfgraph*}(13,10)
				\fmfleft{i}
				\fmfright{o}
				\fmf{gluon}{o,i}
		\end{fmfgraph*}}
		\quad , \quad
		\Delta_{Aa}^{\mu \nu}\equiv
		\parbox{40pt}{
			\begin{fmfgraph*}(13,10)
				\fmfleft{i}
				\fmfright{o}
				\fmf{photon}{i,v}
				\fmf{gluon}{o,v}
		\end{fmfgraph*}}
		\quad , \quad
		\Delta_{\pm \pm}\equiv
		\parbox{40pt}{
			\begin{fmfgraph*}(13,10)
				\fmfleft{i}
				\fmfright{o}
				\fmf{plain}{i,o}
		\end{fmfgraph*}} \quad.
	\end{equation}
\end{fmffile}

Additionally, the Feynman rules for the interaction vertices are:
\begin{fmffile}{vertexrules}\label{vertexrules}
	\begin{equation}
		V_{\pm A^{\mu} \pm} = ie\gamma^{\mu} \equiv \quad
		\parbox{40pt}{
			\begin{fmfgraph*}(15,15)
				\fmfbottom{i,o}
				\fmftop{u}
				\fmf{plain}{i,v}
				\fmf{plain}{o,v}
				\fmf{photon}{v,u}
		\end{fmfgraph*}} \quad \quad , \quad \quad
		V_{\pm a^{\mu} \pm} = \pm ig\gamma^{\mu} \equiv \quad 
		\parbox{40pt}{
			\begin{fmfgraph*}(15,15)
				\fmfbottom{i,o}
				\fmftop{u}
				\fmf{plain}{i,v}
				\fmf{plain}{o,v}
				\fmf{gluon}{v,u}
		\end{fmfgraph*}} \quad \quad .
	\end{equation}
\end{fmffile}

\subsection{The UV and IR Dimensions of the Fields and the Power-Counting}

\quad \,
In order to renormalize the UV divergences, if they exist, the superficial degrees of divergence of the diagrams have to be determined. For each propagator $\D_{XY}$, we can assign an ultraviolet (UV) and an infrared (IR) dimensions $d_{XY}$ and  $r_{XY}$, respectively, given by the propagator behavior on the  UV ($k, ~s \rightarrow \infty$) and IR ($k,~(s-1)\rightarrow 0$) limits. We may define $d_{XY}$ such as $\D_{XY}(k) \sim k^{d_{XY}}$ when $k, ~s \rightarrow \infty$. On the other hand, we have $r_{XY}$ such as $\D_{XY}(k) \sim k^{r_{XY}}$ when $k,~(s-1)\rightarrow 0$. The UV and IR dimensions of the fields,
denoted by $d$ and $r$ respectively, are conditioned by the following inequalities \cite{lz1,lz2}:
\begin{equation}
d_X + d_Y \geq 3 + d_{XY}~,~~~~~~ r_X + r_Y \leq 3 + r_{XY}~, \label{uv-ir}
\end{equation}
For instance, the propagator (\ref{proaa}) in the ultraviolet limit, when $k,~s\rightarrow \infty$, behaves asymptotically as $\displaystyle \D_{AA}^{\m\n}(k)\sim k^{-2}$, while in the infrared limit, when $k,~(s-1)\rightarrow 0$, it behaves as $\displaystyle \D_{AA}^{\m\n}(k)\sim k^0$. Therefore, we can assign the values $d_{AA}=-2$ and $r_{AA}=0$ to the UV and IR dimensions of that propagator, respectively. Thus, applying the inequalities (\ref{uv-ir}) to the $\D_{AA}^{\m\n}(k)$ propagator yields the following relations:
\begin{equation}
	2 d_A\geq 1~,~~~~~ 2r_A \leq 3~.
\end{equation}
The previous procedure can be repeated for the remaining propagators, allowing us to derive the inequalities involving the UV and IR dimensions of all fields. We collect these dimensions in the Table \ref{table1} presented below.
\begin{table}[h!]
	\begin{center}
		\begin{tabular}{|c|c|c|c|c|c|c|c|c|c|c|c|c|}
			\hline
			& $\j_+$ & $\j_-$      & $A_\m$       & $a_\m$ & $b$ & $\pi$ &$c$  &${\ov c}$ &  $\x$    & $\bar \x$     & $s$ & $s-1$  \\
			\hline
			$d$ & 1 & 1 & ${1\over 2}$ & ${1\over 2}$ & ${3\over2}$ & ${3\over 2}$ & 0 & 1& 0 & 1& 1 & 1  \\
			\hline
			$r$ & 1 & 1 &   1          &   1          &   1         &    1         & 0 & 1& 0 & 1& 0 & 1  \\
			\hline
		\end{tabular}
	\end{center}
	\caption[]{UV ($d$) and IR ($r$) dimensions of the fields.}\label{table1}
\end{table}

Taking into account all the previous results, we can determine the power-counting of the model, establishing the superficial degrees of divergence of an arbitrary 1PI diagram $\gamma$:
\begin{equation}
	\bordermatrix{ & \cr & d(\g) \cr
		&r(\g)  }
	= 3 - \sum\limits_f 
	\bordermatrix{& \cr                   
		& d_f \cr
		& r_f  } N_f  -
	\sum\limits_b
	\bordermatrix{ & \cr & d_b \cr
		&\frac{3}{2}r_b  } N_b +
	\bordermatrix{ & \cr & - \cr
		& +  }\frac{1}{2} N_e + 
	\bordermatrix{ & \cr & - \cr
		& +  } \frac{1}{2}N_g -  N_{Aa} ~. \label{power_counting}
\end{equation}
In the power-counting formula, $N_f$ and $N_b$ are the numbers of external fermionic and bosonic lines, respectively, while $N_{Aa}$ is the number of internal lines associated with the mixed propagator\footnote{Despite not carrying any propagating degree of freedom, the mixed propagator can appear when considering all possible virtual processes, that is, they must be taken into account in the Feynman diagrams. It is important to note that $\D_{Aa}^{\m\n}\sim k^{-3}$ in the ultraviolet and $\D_{Aa}^{\m\n}\sim k^{-1}$ in the infrared. In both cases, the momentum power is one unit lower than in the propagators $\D_{AA}^{\m\n}$ and $\D_{aa}^{\m\n}$, which means that diagrams containing internal lines with mixed propagators necessarily have lower superficial degrees of divergence.} $\D_{Aa}^{\m\n}$. Furthermore, $N_e$ and $N_g$ are respectively the power of the coupling constants $e$ and $g$ that appear in the integral representation of each diagram. It is important to note that the model is superrenormalizable since, by increasing the number of vertices, the number of coupling constants naturally increases, which reduces the superficial UV degree of divergence of the diagrams as we increase the loop orders.


\subsection{Feynman diagrams at 1-loop and 2-loops}
The 1-loop diagrams for vacuum polarizations, self-energies and vertex functions are presented in Figure \ref{1loopgraphs}. Additionally, using the power-counting formula (\ref{power_counting}), we present in Table \ref{table2} the superficial degrees of UV and IR divergence for each of these diagrams. It should be mentioned that for any diagram $\gamma_{i_{\pm}}$, the subscript $\pm$ refers to the lines of internal or external legs with $\j_+$ or $\j_-$.

\begin{figure}[H]
	\center
	\begin{fmffile}{vacuumpolarization}
		\begin{fmfgraph*}(30,20)
			\fmfleft{i}
			\fmfv{label=$\gamma_{1_{\pm}}$,label.angle=80,label.dist=1cm}{i}
			\fmfright{o}
			\fmf{photon}{i,v1}
			\fmf{photon}{v2,o}
			\fmf{plain,left,tension=0.4}{v1,v2,v1}
		\end{fmfgraph*}
		\quad \quad 
		\begin{fmfgraph*}(30,20)
			\fmfleft{i}
			\fmfv{label=$\gamma_{2_{\pm}}$,label.angle=80,label.dist=1cm}{i}
			\fmfright{o}
			\fmf{gluon}{v1,i}
			\fmf{gluon}{o,v2}
			\fmf{plain,left,tension=0.4}{v1,v2,v1}
		\end{fmfgraph*}
		\quad \quad
		\begin{fmfgraph*}(30,20)
			\fmfleft{i}
			\fmfv{label=$\gamma_{3_{\pm}}$,label.angle=80,label.dist=1cm}{i}
			\fmfright{o}
			\fmf{photon}{i,v1}
			\fmf{gluon}{o,v2}
			\fmf{plain,left,tension=0.4}{v1,v2,v1}
		\end{fmfgraph*}
		\quad
		\begin{fmfgraph*}(40,20)
			\fmftop{i1}
			\fmfleft{i2,i4}
			\fmfv{label=$\gamma_{4_{\pm}}$,label.angle=80,label.dist=2cm}{i2}
			\fmfright{o,o1}
			\fmf{dashes}{i1,v1}
			\fmf{plain}{i2,v2}
			\fmf{plain}{v3,o}
			\fmf{plain}{v2,v1}
			\fmf{plain}{v3,v1}
			\fmf{dashes}{v2,v3}
		\end{fmfgraph*}	
	\end{fmffile}
\end{figure}

\begin{figure}[H]
	\center
	\begin{fmffile}{SelfenergyAa}
		\begin{fmfgraph*}(30,10)
			\fmfipair{i,b,c,o}
			\fmfiequ{i}{(0,0)}
			\fmfiv{label=$\gamma_{5_{\pm}}$,label.angle=80,label.dist=1cm}{i}
			\fmfiequ{b}{(.3w,0)}
			\fmfiequ{c}{(.7w,0)}
			\fmfiequ{o}{(.9w,0)}
			\fmfi{photon}{c{up} .. tension .9 .. {down}b}
			\fmfi{plain}{i{right} .. {left}b}
			\fmfi{plain}{b{right} .. {left}c}
			\fmfi{plain}{c{right} .. {left}o}
		\end{fmfgraph*}
		\quad \quad
		\begin{fmfgraph*}(30,10)
			\fmfipair{i,b,c,o}
			\fmfiequ{i}{(0,0)}
			\fmfiv{label=$\gamma_{6_{\pm}}$,label.angle=80,label.dist=1cm}{i}
			\fmfiequ{b}{(.3w,0)}
			\fmfiequ{c}{(.7w,0)}
			\fmfiequ{o}{(.95w,0)}
			\fmfi{gluon}{c{up} .. tension 1 .. {down}b}
			\fmfi{plain}{i{right} .. {left}b}
			\fmfi{plain}{b{right} .. {left}c}
			\fmfi{plain}{c{right} .. {left}o}
		\end{fmfgraph*}
		\quad \quad
		\begin{fmfgraph*}(30,10)
			\fmfipair{i,b,a,c,o}
			\fmfiequ{i}{(0,0)}
			\fmfiv{label=$\gamma_{7_{\pm}}$,label.angle=80,label.dist=1cm}{i}
			\fmfiequ{b}{(.3w,0)}
			\fmfiequ{c}{(.7w,0)}
			\fmfiequ{a}{(.5w,0.8cm)}
			\fmfiequ{o}{(.95w,0)}
			\fmfi{photon}{b{up} .. tension 1 .. {right}a}
			\fmfi{gluon}{c{up} .. tension 1 .. {left}a}
			\fmfi{plain}{i{right} .. {left}b}
			\fmfi{plain}{b{right} .. {left}c}
			\fmfi{plain}{c{right} .. {left}o}
		\end{fmfgraph*}
	\end{fmffile}
	\caption{The 1-loop diagrams $\gamma_{1_{\pm}}$, $\gamma_{2_{\pm}}$, and $\gamma_{3_{\pm}}$ represent the vacuum polarization tensors, $\gamma_{4_{\pm}}$ represents the interaction vertices, and $\gamma_{5_{\pm}}$, $\gamma_{6_{\pm}}$, and $\gamma_{7_{\pm}}$ are the fermion self-energies. The solid lines represent external legs or the propagators of $\j_+$ or $\j_-$, while the dashed lines in $\gamma_{4_{\pm}}$ denote propagators of $A_\m$, $a_\m$, the mixed propagator or external legs of $A_\m$ or $a_\m$.}\label{1loopgraphs}
\end{figure}
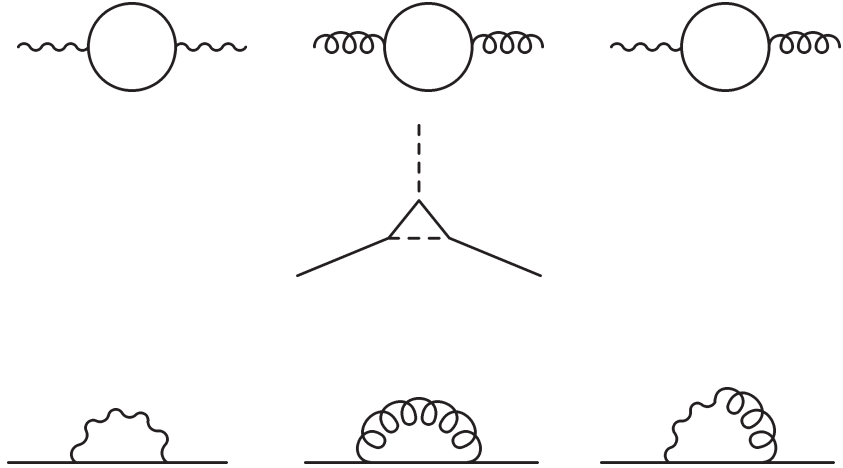

\begin{table}[h]
	\begin{center}
		\begin{tabular}{|c|c|c|c|c|c|c|c|c|}
			\hline
			&$\gamma_{1_{\pm}}$ & $\gamma_{2_{\pm}}$ & $\gamma_{3_{\pm}}$           & $\gamma_{4_{\pm}}^{(a)}$ & $\gamma_{4_{\pm}}^{(b)}$    &  $\gamma_{5_{\pm}}$   & $\gamma_{6_{\pm}}$  &$\gamma_{7_{\pm}}$    \\
			\hline
			$d$      & 1     &    1   &   1   &  $-1$  & $-2$  & 0   &  0 &  $-1$      \\
			\hline
			$r$      &  1    &   1    &    1             &   1      &  0    &   2           &    2         &  1        \\
			\hline
		\end{tabular}
	\end{center}
	\caption[]{
			Superficial degrees of divergence of the diagrams in Figure \ref{1loopgraphs}.
			$\g_{4\pm}^{(b)}$ is obtained when considering the internal dashed line as a mixed propagator, whereas $\g_{4\pm}^{(a)}$ is obtained when considering it as $\D_{AA}$ or $\D_{aa}$.
	} \label{table2}
\end{table}

According to the Power-Counting Theorem, $\g_{4_\pm}$ and $\g_{7_\pm}$ are absolutely convergent, since they possess $d$ lower than zero. On the other hand, the diagrams 
$\gamma_{1_{\pm}}$, $\gamma_{2_{\pm}}$, $\gamma_{3_{\pm}}$, $\gamma_{5_{\pm}}$ and $\gamma_{6_{\pm}}$ are superficially divergent in the UV limit, and therefore must be subjected to some renormalization procedure. For this purpose, the BPHZL subtraction scheme will be used and it will be discussed in the following section.

At 2-loops order, we have the following vacuum polarization diagrams:
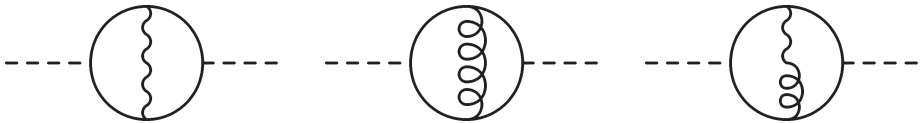
\begin{figure}[H]
	\center
	\begin{fmffile}{twoloopgraphs}
		\begin{fmfgraph*}(37,25)
			\fmfipair{i,va,vb,vc,vd,o}
			\fmfiequ{i}{(0,.5h)}
			\fmfiv{label=$\gamma_{8_{\pm}}$,label.angle=80,label.dist=1cm}{i}
			\fmfiequ{va}{(.3w,.5h)}
			\fmfiequ{vb}{(.5w,.8h)}
			\fmfiequ{vc}{(.7w,.5h)}
			\fmfiequ{vd}{(.5w,.2h)}
			\fmfiequ{o}{(w,.5h)}
			\fmfi{dashes}{i--va}
			\fmfi{dashes}{vc{right} .. {right}o}
			\fmfi{plain}{va{up} .. tension 1 .. {right}vb}
			\fmfi{plain}{vb{right} .. tension 1 .. {down}vc}
			\fmfi{plain}{vc{down} .. tension 1 .. {left}vd}
			\fmfi{plain}{vd{left} .. tension 1 .. {up}va}
			\fmfi{photon}{vb{down} .. {down}vd}
		\end{fmfgraph*}
		\quad
		\begin{fmfgraph*}(37,25)
			\fmfipair{i,va,vb,vc,vd,o}
			\fmfiequ{i}{(0,.5h)}
			\fmfiv{label=$\gamma_{9_{\pm}}$,label.angle=80,label.dist=1cm}{i}
			\fmfiequ{va}{(.3w,.5h)}
			\fmfiequ{vb}{(.5w,.8h)}
			\fmfiequ{vc}{(.7w,.5h)}
			\fmfiequ{vd}{(.5w,.2h)}
			\fmfiequ{o}{(w,.5h)}
			\fmfi{dashes}{i--va}
			\fmfi{dashes}{vc{right} .. {right}o}
			\fmfi{plain}{va{up} .. tension 1 .. {right}vb}
			\fmfi{plain}{vb{right} .. tension 1 .. {down}vc}
			\fmfi{plain}{vc{down} .. tension 1 .. {left}vd}
			\fmfi{plain}{vd{left} .. tension 1 .. {up}va}
			\fmfi{gluon}{vd{up} .. {up}vb}
		\end{fmfgraph*}
		\quad
		\begin{fmfgraph*}(37,25)
			\fmfipair{i,va,vb,vc,vd,o,c}		
			\fmfiequ{i}{(0,.5h)}
			\fmfiv{label=$\gamma_{10_{\pm}}$,label.angle=80,label.dist=1cm}{i}
			\fmfiequ{va}{(.3w,.5h)}
			\fmfiequ{vb}{(.5w,.8h)}
			\fmfiequ{vc}{(.7w,.5h)}
			\fmfiequ{vd}{(.5w,.2h)}
			\fmfiequ{o}{(w,.5h)}
			\fmfiequ{c}{(.5w,.5h)}
			\fmfi{dashes}{i--va}
			\fmfi{dashes}{vc{right} .. {right}o}
			\fmfi{plain}{va{up} .. tension 1 .. {right}vb}
			\fmfi{plain}{vb{right} .. tension 1 .. {down}vc}
			\fmfi{plain}{vc{down} .. tension 1 .. {left}vd}
			\fmfi{plain}{vd{left} .. tension 1 .. {up}va}
			\fmfi{photon}{vb{down} .. {down}c}
			\fmfi{gluon}{vd{up} .. {up}c}
		\end{fmfgraph*}
	\end{fmffile} 
\end{figure}
\begin{figure}[H]
	\center
	\begin{fmffile}{2loops}
		\begin{fmfgraph*}(37,25)
			\fmfipair{i,va,vb,vc,vd,o,c,ve,vf}		
			\fmfiequ{i}{(0,.5h)}
			\fmfiv{label=$\gamma_{11_{\pm}}$,label.angle=80,label.dist=1cm}{i}
			\fmfiequ{va}{(.3w,.5h)}
			\fmfiequ{vb}{(.5w,.8h)}
			\fmfiequ{vc}{(.7w,.5h)}
			\fmfiequ{vd}{(.5w,.2h)}
			\fmfiequ{o}{(w,.5h)}
			\fmfiequ{c}{(.5w,.5h)}
			\fmfiequ{ve}{c+(-.17w,-.15h)}
			\fmfiequ{vf}{c+(.17w,-.15h)}
			\fmfi{photon}{ve{up} .. tension 2.3 .. {down}vf}
			\fmfi{dashes}{i--va}
			\fmfi{dashes}{vc{right} .. {right}o}
			\fmfi{plain}{va{up} .. tension 1 .. {right}vb}
			\fmfi{plain}{vb{right} .. tension 1 .. {down}vc}
			\fmfi{plain}{vc{down} .. tension 1 .. {left}vd}
			\fmfi{plain}{vd{left} .. tension 1 .. {up}va}
		\end{fmfgraph*}	
		\quad
		\begin{fmfgraph*}(37,25)
			\fmfipair{i,va,vb,vc,vd,o,c,ve,vf}		
			\fmfiequ{i}{(0,.5h)}
			\fmfiv{label=$\gamma_{12_{\pm}}$,label.angle=80,label.dist=1cm}{i}
			\fmfiequ{va}{(.3w,.5h)}
			\fmfiequ{vb}{(.5w,.8h)}
			\fmfiequ{vc}{(.7w,.5h)}
			\fmfiequ{vd}{(.5w,.2h)}
			\fmfiequ{o}{(w,.5h)}
			\fmfiequ{c}{(.5w,.5h)}
			\fmfiequ{ve}{c+(-.17w,-.15h)}
			\fmfiequ{vf}{c+(.17w,-.15h)}
			\fmfi{curly}{vf .. tension 1 .. ve}
			\fmfi{dashes}{i--va}
			\fmfi{dashes}{vc{right} .. {right}o}
			\fmfi{plain}{va{up} .. tension 1 .. {right}vb}
			\fmfi{plain}{vb{right} .. tension 1 .. {down}vc}
			\fmfi{plain}{vc{down} .. tension 1 .. {left}vd}
			\fmfi{plain}{vd{left} .. tension 1 .. {up}va}
		\end{fmfgraph*}	
		\quad
		\begin{fmfgraph*}(37,25)
			\fmfipair{i,va,vb,vc,vd,o,c,ve,vf}		
			\fmfiequ{i}{(0,.5h)}
			\fmfiv{label=$\gamma_{13_{\pm}}$,label.angle=80,label.dist=1cm}{i}
			\fmfiequ{va}{(.3w,.5h)}
			\fmfiequ{vb}{(.5w,.8h)}
			\fmfiequ{vc}{(.7w,.5h)}
			\fmfiequ{vd}{(.5w,.2h)}
			\fmfiequ{o}{(w,.5h)}
			\fmfiequ{c}{(.5w,.5h)}
			\fmfiequ{ve}{c+(-.17w,-.15h)}
			\fmfiequ{vf}{c+(.17w,-.15h)}
			\fmfi{curly}{vf--c}
			\fmfi{photon}{c{left} .. tension .8 .. ve}
			\fmfi{dashes}{i--va}
			\fmfi{dashes}{vc{right} .. {right}o}
			\fmfi{plain}{va{up} .. tension 1 .. {right}vb}
			\fmfi{plain}{vb{right} .. tension 1 .. {down}vc}
			\fmfi{plain}{vc{down} .. tension 1 .. {left}vd}
			\fmfi{plain}{vd{left} .. tension 1 .. {up}va}
		\end{fmfgraph*}	
		\caption{Vacumm polarizations at 2-loops, where the continuous lines represent propagators of $\j_+$ or of $\j_-$, while dashed lines represent external legs of the fields $A_\m$ or $a_\m$.}\label{2loopgraphs}
	\end{fmffile}
\end{figure}
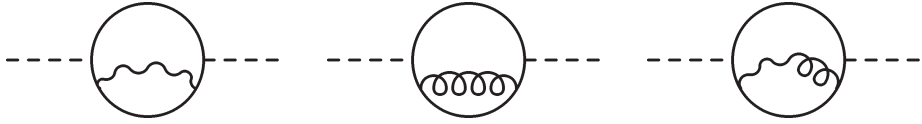
There are 36 types of vacuum polarization diagrams, with 12 of them being convergent, namely those containing mixed propagators as internal lines, which are represented by $\gamma_{10_\pm}$ and $\gamma_{13_\pm}$, with superficial UV degrees of divergence are $d(\gamma_{10_\pm})=d(\gamma_{13_\pm})=-1$. In contrast, $\gamma_{8_\pm}$, $\gamma_{9_\pm}$, $\gamma_{11_\pm}$, and $\gamma_{12_\pm}$ are potentially logarithmically divergent since $d(\gamma_{8_{\pm}})=d(\gamma_{9_{\pm}})=d(\gamma_{11_{\pm}})=d(\gamma_{12_{\pm}})=0$ and need to be renormalized. In addition, these are the last diagrams that need a renormalization procedure, as can be checked by the power counting formula, since from 3-loops onward all diagrams have $d<0$. Therefore, to complete the renormalization of the model, namely its ultraviolet and infrared subtractions, we must go up to loop order 2.

\section{The BPHZL Subtraction Scheme}
\quad
When dealing with massless fields, an appropriate renormalization procedure is the BPHZL subtraction scheme. This is because of the need to handle the IR divergences that arise due to the renormalization of UV divergences. This is the reason for choosing BPHZL over, for example, the BPHZ \cite{BPHZ} (Bogoliubov-Parasiuk-Hepp-Zimmermann) subtraction scheme. Before presenting the BPHZL method, some definitions need to be introduced.
For further details, see \cite{lz1,lz2} 

\begin{mydef}
	Let $\Lambda$ be a Feynman diagram. A subdiagram $\l_i$ of $\Lambda$ is a subset of vertices within $\Lambda$ along with the lines that connect them. A subdiagram is called a trivial subdiagram if it either consists of the empty diagram or represents the entire diagram itself. Conversely, a subdiagram is called a proper diagram if it is not trivial.
	
\end{mydef}
From now on, a subdiagram has to be intended as a 1PI subdiagram.

\begin{mydef}
	Consider two subdiagrams $\l_i$ and $\l_j$. Also, consider the following conditions:
	\begin{equation}
		\l_i \subseteq \l_j~;~~~\l_j \subseteq \l_i~;~~~\l_i \cap \l_j=\varnothing~.
	\end{equation}
	If none of these conditions is satisfied, we say that $\l_i$ and $\l_j$ overlap. If $\l_i \cap \l_j=\varnothing$, we say that $\l_i$ and $\l_j$ are disjoint.
\end{mydef}

\begin{mydef}
	Let $\{\l_1\cdots\l_n\}$ be a set of mutually disjoint proper subdiagrams. A reduced subdiagram $\Lambda/\l_1\cdots\l_n$ is obtained from $\Lambda$ by contracting all $\l_i,~i=1,\cdots,n$, to points (reduced vertices). Under these conditions, it follows that
	\begin{align}
		& d(\Lambda)=d(\Lambda/\l_1\cdots\l_n)+\sum_{i=1}^{n}d(\l_i)~,\\
		& r(\Lambda)=r(\Lambda/\l_1\cdots\l_n)+\sum_{i=1}^{n}r(\l_i)~.
	\end{align}
\end{mydef}

\begin{mydef}
	Let $U(\Lambda)$ be a set of subdiagrams, proper or trivial, of a diagram $\Lambda$. Let $\l_i,\l_j \subseteq \Lambda$ be two arbitrary elements of $U(\Lambda)$. If $\l_i,\l_j$ do not overlap, for
	all $\l_i,\l_j \in U(\Lambda)$, then we say that $U(\Lambda)$ is a $\Lambda$-forest. 
\end{mydef}

\begin{mydef}
	Given a subdiagram $\l$ of a diagram $\Lambda$, consider the UV and IR subtraction degrees, denoted by $\d(\l)$ and $\r(\l)$, respectively. Let $\{\l_1,\cdots,\l_n\}$ be an arbitrary set of subdiagrams of $\l$, and $\l/\l_1\cdots\l_n$ be an arbitrary reduced diagram. Under these conditions, the following inequalities hold:
	\begin{subequations}
		\begin{equation}\label{consistencyb}
			\d(\l)\geq d(\l)+b(\l)~,
		\end{equation}
		\begin{equation}\label{consistencyc}
			\r({\l}) \leq r(\l)-c(\l)~,
		\end{equation}
	\end{subequations}
	where $b(\l)$ and $c(\l)$ are positive integers satisfying the following constraints:
	\begin{subequations}\label{vinc}
		\begin{equation}
			\d(\l)\geq d(\l/\l_1\cdots\l_n)+\sum_{i=1}^n \d(\l_i)~,
		\end{equation}
		\begin{equation}
			\r(\l)\leq r(\l/\l_1\cdots\l_n)+\sum_{i=1}^n \r(\l_i)~,
		\end{equation}
		\begin{equation}
			\r(\l)\leq \d(\l)+1~.~~~~~~~~~~~~~~\,\,\,\,\,\,\,\,\,\,\,~~
		\end{equation}
	\end{subequations}
\end{mydef}

Now we are in a position to define how to subtract the UV divergences of a divergent diagram using the BPHZL procedure.
Consider an arbitrary $n$-loop diagram $\Lambda$, divergent in the ultraviolet limit, with the integral given by
\begin{equation}
 \Lambda = \int d^3 k_1 \cdots d^3 k_n I_\Lambda(p,k,s)~,
\end{equation}
where $k_i$ represent internal momenta, $p$ represents external momenta, and $s$ is the Lowenstein-Zimmermann parameter. The renormalized integrand $R_\Lambda(p,k,s)$, is given by Zimmermann's forest formula:
\begin{equation}\label{forest}
	R_\Lambda(p,k,s)= S_\Lambda \sum_{U\in \mathcal{F}_\Lambda}\prod_{\l\in U}(-\t_\l S_\l )I_\Lambda(p,k,s)~.
\end{equation}
Here, $S_\g$ is the substitution operator, which alters the $p$, $k$, and $s$, in the lines corresponding to the subdiagram $\g$, to $p^\g$, $k^\g$, and $s^\g$, while $\mathcal{F}_\Lambda$ is the family of all the forests of $\Lambda$. Additionally, $\t_\lambda$ is the operator that acts in the terms corresponding to $\g$, defined by
\begin{equation}
	1-\t_\l=\left(1-t^{\r(\l)-1}_{p^\l,(s^\l-1)}\right)
	\left(1-t^{\d(\l)}_{p^\l,s^\l}\right)~,
\end{equation}
where
\begin{equation}
	t^d_{x,y}= 
	\begin{cases}
		&\textrm{Taylor polynomial of degree $d$ around}~ x= y=0 \textrm{ if } d\geq 0 \\
		& 0 \textrm{ if } d<0
	\end{cases}~.
\end{equation}
Furthermore, we can define $-\t_\l=1$ if $\l=\varnothing$. Another important point to mention is that if a forest contains $\l_i$, $\l_j$, such that $\l_i\subset \l_j$, then, in the product in (\ref{forest}), we must start from the smaller diagram to the larger, that is, $(-\t_{\l_i}S_{\l_i})$ should act first on $I_{\Lambda}(p,k,s)$ than $(-\t_{\l_j}S_{\l_j})$. We emphasize that $\varnothing$ is also considered a forest. Finally, it is important to note that forests containing at least one convergent subdiagram do not contribute to (\ref{forest}). Hence, we take into consideration forests that contain only divergent subdiagrams, which are called the renormalization parts of $\Lambda$. Once the subtraction procedure is performed, the limit $s=1$ has to be taken, recovering the massless case.

An important fact to note is that 1-loop diagrams have no proper subdiagrams. In
this case, the renormalized integrand is just
\begin{equation}\label{R}
	R_{\gamma}(p,k,s)=\left(1-t^{\rho(\gamma)-1}_{p,s-1}\right)
	\left(1-t^{\delta(\gamma)}_{p,s}\right)I_{\gamma}(p,k,s) ~,
\end{equation}
and the constraints (\ref{vinc}) become
\begin{align}\label{redvinc}
	\nonumber
	\d(\g)=d(\g)+b(\g)~,~~~\r(\g)=r(\g)-c(\g)~;\\
	\d(\g)\geq d(\g)~,~~~\r(\g)\leq r(\g)~,~~~\r(\g)\leq \d(\g)+1~&.
\end{align}
The 1-loop vacuum polarization renormalized diagrams, presented in Figure \ref{1loopgraphs}, are
\begin{equation}
	\Pi^{(R)\m\n}_{\g_{1\pm}}(p,1)=-\frac{e^2}{16}\frac{\eta^{\m\n}p^2-p^\m p^\n}{\sqrt{p^2}}\mp\frac{e^2 m}{4\pi |m|}\ve^{\m\a\n}p_\a~.
\end{equation}
\begin{equation}
	\Pi^{(R)\m\n}_{\g_{2\pm}}(p,1)=-\frac{g^2}{16}\frac{\eta^{\m\n}p^2-p^\m p^\n}{\sqrt{p^2}}\mp\frac{g^2 m}{4\pi |m|}\ve^{\m\a\n}p_\a~,
\end{equation}
\begin{equation}
	\Pi^{(R)\m\n}_{\g_{3\pm}}(p,1)=\frac{eg}{16}\frac{\eta^{\m\n}p^2-p^\m p^\n}{\sqrt{p^2}}\pm\frac{eg m}{4\pi |m|}\ve^{\m\a\n}p_\a~,
\end{equation}
and for the self-energies we have
\begin{dmath}
	\S^{(R)}_{\gamma_{5_{\pm}}}
	=-\frac{ie^2 {\Sl{p}}}{4\p}\left[\frac{1}{4\sqrt{p^2}}\left(\frac{p^2}{\m^2}+
	\frac{3\m^2}{p^2}+2\right)
	\ln \left(\frac{\m^2-p^2}{(\sqrt{\m^2}-\sqrt{p^2})^2}\right)
	-\frac{|\m|}{2}\left(\frac{1}{\m^2}+\frac{3}{p^2}\right)
	+i\p\frac{p^2}{4\m^2\sqrt{p^2}}
	\right]~,
\end{dmath}
\begin{dmath}
	\S^{(R)}_{\gamma_{6_{\pm}}}
	=-\frac{ig^2 \Sl{p}}{4\p} \left[\frac{1}{4\sqrt{p^2}}\left(\frac{p^2}{\m^2}+
	\frac{3\m^2}{p^2}+2\right)
	\ln \left(\frac{\m^2-p^2}{(\sqrt{\m^2}-\sqrt{p^2})^2}\right)
	-\frac{|\m|}{2}\left(\frac{1}{\m^2}+\frac{3}{p^2}\right)
	+i\p\frac{p^2}{4\m^2\sqrt{p^2}}
	\right]~.
\end{dmath}

The previous results about 1-loop vacuum polarizations for the gauge fields and self-energies for the fermion fields were presented and discussed in detail in \cite{parity}.

\section{BPHZL at 2-loops}
Before applying the BPHZL to the divergent diagrams at 2-loops, it is necessary to write their expressions. Taking into account Figure \ref{2loopgraphs}, it is possible to see that
\begin{align}  
&\P_{\gamma_{8_{\pm}}}^{\m\n}(p,s)=\l_8^2\int{\frac{d^3k_1}{(2\p)^3}}\int{\frac{d^3k_2}{(2\p)^3}}~e^2~{\wh I}_{\pm}^{\m\n}(k_1,k_2,p,s)~,\label{I8} \\
&\P_{\gamma_{9_{\pm}}}^{\m\n}(p,s)=\l_9^2\int{\frac{d^3k_1}{(2\p)^3}}\int{\frac{d^3k_2}{(2\p)^3}}~g^2~{\wh I}_{\pm}^{\m\n}(k_1,k_2,p,s)~;\label{I9}
\end{align}
and 
\begin{align}  
&\P_{\gamma_{11_{\pm}}}^{\m\n}(p,s)=\l_{11}^2\int{\frac{d^3k_1}{(2\p)^3}}\int{\frac{d^3k_2}{(2\p)^3}}~e^2~{\wt I}_{\pm}^{\m\n}(k_1,k_2,p,s)~,\label{I11} \\
&\P_{\gamma_{12_{\pm}}}^{\m\n}(p,s)=\l_{12}^2\int{\frac{d^3k_1}{(2\p)^3}}\int{\frac{d^3k_2}{(2\p)^3}}~g^2~{\wt I}_{\pm}^{\m\n}(k_1,k_2,p,s)~;\label{I12}
\end{align}
where $\l_i=e$ ($i=8,9,11,12$) if the two external legs are $A_\m$ and, if they are then $a_\m$, $\l_i=g$. Furthermore, 
\begin{align}
	\nonumber
	{\wh I}_{\pm}^{\m\n}&(k_1,k_2, p,s)=-{\rm Tr}\left\{
	\gamma^{\mu} 
	\left[ i \frac{\Sl k_1\mp m(s-1)}{k_1^2-m^2(s-1)^2} 
	\right]
	\gamma_{\alpha}
	\left[-i\frac{1}{(k_1-k_2)^2-\mu^2}\times \right. \right. \\ \nonumber \\
	\nonumber
	&\times \left.\left(\eta^{\alpha \beta}-\frac{(k_1^{\alpha}-k_2^{\alpha})(k_1^{\beta}-
		k_2^{\beta})}{(k_1-k_2)^2}\right)\right]
	\left.\left[ i \frac{\Sl k_2\mp m(s-1)}{k_2^2-m^2(s-1)^2} \right]
	\gamma^{\nu}\times \right. \\ \nonumber \\
	&\left.\times
	\left[ i \frac{(\Sl k_2- \Sl p)\mp m(s-1)}{(k_2-p)^2-m^2(s-1)^2} \right]
	\gamma_{\beta}	
	\left[ i \frac{(\Sl k_1- \Sl p)\mp m(s-1)}{(k_1-p)^2-m^2(s-1)^2} \right]	
	\right\}~, \label{I-tilde}
\end{align}
whereas

\begin{align}
	\nonumber
	{\wt I}_{\pm}^{\m\n}(k_1,& k_2,p,s)=-{\rm Tr}\left\{
	\gamma^{\mu} 
	\left[ i \frac{\Sl k_1\mp m(s-1)}{k_1^2-m^2(s-1)^2} 
	\right]
	\gamma^{\n}
	\left[ i \frac{(\Sl k_1- \Sl p)\mp m(s-1)}{(k_1-p)^2-m^2(s-1)^2} \right] \right. \times \\ \nonumber \\
	\nonumber
	&\times\g_\a 
	\left[-i\frac{1}{k_2^2-\mu^2}
	\left(\eta^{\alpha \beta}-\frac{k_2^{\alpha}k_2^{\beta}}
	{k_2^2}\right)\right] 
	\left[ i \frac{(\Sl k_1-\Sl k_2-\Sl p)\mp m(s-1)}{(k_1-k_2-p)^2-m^2(s-1)^2} \right] \times \\ \nonumber \\
	&~~~~~~~~~~~~~~~~~~~~~~~~~~~~~~~~~\times\gamma_{\b}\left.	
	\left[ i \frac{(\Sl k_1- \Sl p)\mp m(s-1)}{(k_1-p)^2-m^2(s-1)^2} \right]
	\right\}~.\label{I-hat}
\end{align}

\subsection{The Zimmermann's Forests}
\quad \,
As discussed previously, the diagrams shown in Figure \ref{2loopgraphs} that require renormalization are $\gamma_{8_\pm}$, $\gamma_{9_\pm}$, $\gamma_{11_\pm}$, and $\gamma_{12_\pm}$. The first step in applying the BPHZL subtraction scheme at the 2-loop level is to identify the subdiagrams and determine whether they are divergent, indicating the presence of subdivergences within the complete diagram. It is also necessary to check whether the subdiagrams overlap to construct the Zimmermann's forests. Notably,the diagrams $ \gamma_{8_\pm}$ and $\gamma_{9_\pm}$ are topologically equivalent, and in the Landau gauge, their differences stem solely from distinct coupling constants derived from the Feynman rules. Similar conclusions can be drawn for the diagrams $\gamma_{11_\pm}$ and $\gamma_{12_\pm}$.

Let us start analyzing the subdiagrams originated from $\g_{8_\pm}$. In the following figure, all the proper subdiagrams of $\g_{8_\pm}$ are presented. 
\vspace{1cm}
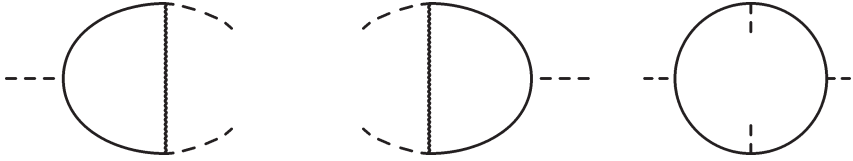
\begin{figure}[H]\label{sub8}
	\center
	\begin{fmffile}{sub8} 
		\begin{fmfgraph*}(30,20)
			\fmfset{wiggly_len}{15}
			\fmfipair{i,oa,ob,vi,va,vb}
			\fmfiequ{i}{(0,.5h)}
			\fmfiequ{oa}{(w,.8h)}
			\fmfiequ{ob}{(w,.2h)}
			\fmfiequ{vi}{(.25w,.5h)}	
			\fmfiequ{va}{(.7w,h)}
			\fmfiequ{vb}{(.7w,0)}
			\fmfi{dashes}{i--vi}
			\fmfi{plain}{vi{up} .. tension 1 .. {right}va}
			\fmfi{plain}{vi{down} .. tension 1 .. {right}vb}
			\fmfi{wiggly}{va--vb}
			\fmfi{dashes}{va{right} .. tension 2 .. {down}oa}
			\fmfi{dashes}{vb{right} .. tension 2 .. {up}ob}
			\fmfiv{label=$\g_8'$,label.angle=90,label.dist=1.3cm}{i}
		\end{fmfgraph*}
		\quad \quad \quad \quad 
		\begin{fmfgraph*}(30,20)
			\fmfset{wiggly_len}{15}
			\fmfipair{i,oa,ob,vi,va,vb}
			\fmfiequ{i}{(w,.5h)}
			\fmfiequ{oa}{(0,.8h)}
			\fmfiequ{ob}{(0,.2h)}
			\fmfiequ{vi}{(.75w,.5h)}	
			\fmfiequ{va}{(.3w,h)}
			\fmfiequ{vb}{(.3w,0)}
			\fmfi{dashes}{i--vi}
			\fmfi{plain}{vi{up} .. tension 1 .. {left}va}
			\fmfi{plain}{vi{down} .. tension 1 .. {left}vb}
			\fmfi{wiggly}{va--vb}
			\fmfi{dashes}{va{left} .. tension 2 .. {down}oa}
			\fmfi{dashes}{vb{left} .. tension 2 .. {up}ob}
			\fmfiv{label=$\g_8''$,label.angle=90,label.dist=0.8cm}{oa}
		\end{fmfgraph*}
		\begin{fmfgraph*}(40,20)
			\fmfipair{i,vi,va,vb,vo,o,ha,hb}
			\fmfiequ{i}{(0.15w,.5h)}
			\fmfiequ{o}{(0.85w,.5h)}
			\fmfiequ{vo}{(.75w,.5h)}
			\fmfiequ{vi}{(.25w,.5h)}	
			\fmfiequ{va}{(.5w,h)}
			\fmfiequ{vb}{(.5w,0)}
			\fmfiequ{ha}{(.5w,.75h)}
			\fmfiequ{hb}{(.5w,.25h)}
			\fmfi{dashes}{i--vi}
			\fmfi{plain}{vi{up} .. tension 1 .. {right}va}
			\fmfi{plain}{vi{down} .. tension 1 .. {right}vb}
			\fmfi{dashes}{va--ha}
			\fmfi{dashes}{vb--hb}
			\fmfi{plain}{va{right} .. tension 1 .. {down}vo}
			\fmfi{plain}{vb{right} .. tension 1 .. {up}vo}
			\fmfi{dashes}{vo--o}
			\fmfiv{label=$\g_8'''$,label.angle=90,label.dist=1.3cm}{i}
		\end{fmfgraph*} 
		\quad \quad \quad \quad \quad \quad \quad \quad \quad \quad 
	\end{fmffile}
	\caption{1PI subdiagrams originated from $\g_{8_\pm}$. Dashed lines must
		be treated as external lines for these subdiagrams, even though some
		them are internal lines for the complete diagram.}	
\end{figure}
\noindent
As can be seen, $\gamma_{8}'$ and ${\gamma_{8}''}$ are 1-loop vertex-type diagrams, which are convergent (see Table \ref{1loopgraphs}). Additionally, $\gamma_8'''$ is also convergent. Thus, there are no subdivergences for $\gamma_{8}$, and the BPHZL subtraction will be entirely analogous to that performed at the 1-loop. This becomes evident when we determine the family of forests for $\gamma_{8}$:
\begin{equation}\label{forest1}
	\mathcal{F}(\g_{8_\pm})=\{\varnothing, \{\g_{8_\pm}\}\}~,
\end{equation}
Besides, as $\g_{9_\pm}$ is topologically equivalent to $\g_{8_\pm}$, it follows
that the family of forests for this diagram is
\begin{equation}\label{forest2}
	\mathcal{F}(\g_{9_\pm})=\{\varnothing, \{\g_{9_\pm}\}\}~,
\end{equation}

Let us repeat the previous approach, but now considering $\g_{11\pm}$. In this
context, the subdiagrams are represented in the following figure.
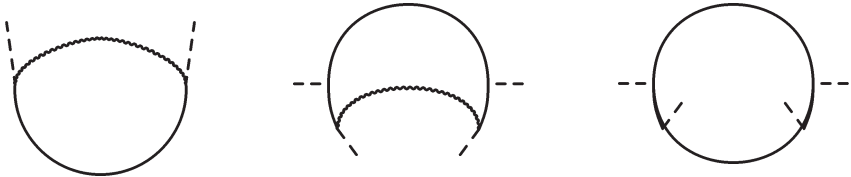
\begin{figure}[H]\label{subd11}
	\center
	\begin{fmffile}{subd11} 
		\vspace{0.4cm}
		\begin{fmfgraph*}(25,30)
			\fmfset{wiggly_len}{17}
			\fmfipair{i,o,va,vb,vi,vo}
			\fmfiequ{i}{(.05w,.5h)}
			\fmfiequ{o}{(0.95w,.5h)}
			\fmfiequ{va}{(0.5w,.7h)}
			\fmfiequ{vb}{(0.5w,.1h)}
			\fmfiequ{vi}{(0,.8h)}
			\fmfiequ{vo}{(w,.8h)}
			\fmfi{plain}{i.. tension 1 .. {right}vb}
			\fmfi{plain}{vb{right}.. tension 1 .. o}
			\fmfi{dashes}{i.. tension 1.5 .. vi}
			\fmfi{dashes}{o.. tension 1.5 .. vo}
			\fmfi{wiggly}{i{up}.. tension 1.5 .. {right}va}
			\fmfi{wiggly}{va{right}.. tension 1.5 .. {down}o}
			\fmfiv{label=$\g_{11}'$,label.angle=90,label.dist=1.3cm}{i}
		\end{fmfgraph*}
		\quad \quad \quad 
		\begin{fmfgraph*}(30,30)
			\vspace{0.4cm}
			\fmfset{wiggly_len}{18}
			\fmfipair{i,o,vi,vo,vii,voo,viii,vooo}
			\fmfiequ{i}{(0,.5h)}
			\fmfiequ{viii}{(.3w,.15h)}
			\fmfiequ{vooo}{(.7w,.15h)}
			\fmfiequ{o}{(w,.5h)}
			\fmfiequ{vi}{(0.15w,.5h)}
			\fmfiequ{vo}{(0.85w,.5h)}
			\fmfiequ{vii}{(0.19w,.3h)}
			\fmfiequ{voo}{(0.81w,.3h)}
			\fmfi{dashes}{i--vi}
			\fmfi{dashes}{o--vo}
			\fmfi{plain}{vi{up} .. tension 1 .. {down}vo}
			\fmfi{plain}{vi{down} .. tension .8 .. vii}
			\fmfi{plain}{vo{down} .. tension .8 .. voo}
			\fmfi{wiggly}{vii{up} .. tension 1.7 .. {down}voo}
			\fmfi{dashes}{vii--viii}
			\fmfi{dashes}{voo--vooo}
			\fmfiv{label=$\g_{11}''$,label.angle=90,label.dist=1.2cm}{i}
		\end{fmfgraph*} 
\quad \quad \quad
		\begin{fmfgraph*}(30,30)
			\fmfipair{i,o,vi,vo,vii,voo,viii,vooo}
			\fmfiequ{i}{(0,.5h)}
			\fmfiequ{viii}{(.3w,.45h)}
			\fmfiequ{vooo}{(.7w,.45h)}
			\fmfiequ{o}{(w,.5h)}
			\fmfiequ{vi}{(0.15w,.5h)}
			\fmfiequ{vo}{(0.85w,.5h)}
			\fmfiequ{vii}{(0.19w,.3h)}
			\fmfiequ{voo}{(0.81w,.3h)}
			\fmfi{dashes}{i--vi}
			\fmfi{dashes}{o--vo}
			\fmfi{plain}{vi{up} .. tension 1 .. {down}vo}
			\fmfi{plain}{vi{down} .. tension 1 .. {up}vo}
			\fmfi{plain}{vi{down} .. tension .8 .. vii}
			\fmfi{plain}{vo{down} .. tension .8 .. voo}
			\fmfi{dashes}{vii--viii}
			\fmfi{dashes}{voo--vooo}
			\fmfiv{label=$\g_{11}'''$,label.angle=90,label.dist=1.2cm}{i}
		\end{fmfgraph*}
	\end{fmffile}
	\caption{
		1PI subdiagrams related to the diagrams $\gamma_{11_\pm}$. Dashed lines must be treated as external lines in these subdiagrams, even though some of them are internal lines when considering the complete diagram.}	
\end{figure}
\noindent
Now the situation is slightly different. The subdiagram $\g_{11}'$ is a self-energy diagram at 1-loop and is divergent, as verified in Table \ref{1loopgraphs}. On the other hand, the subdiagrams $\g_{11}''$ and $\g_{11}'''$ are similar to $\g_{8}'''$, all of them  corresponding to four-point amplitudes of gauge bosons, which are convergent. Therefore, the family of forests for the diagram $\g_{11_\pm}$ is
\begin{equation}\label{forest11}
	\mathcal{F}(\g_{11_\pm})=\{\varnothing, \{\g_{11_\pm}'\},\{\g_{11_\pm}\},\{\g_{11_\pm}',\g_{11_\pm}\}\}~.
\end{equation}
Considering $\g_{12_\pm}$ and denoting the self-energy-like subdiagram as $\g_{12_\pm}'$, the family of forests in this case is given by
\begin{equation}\label{forest12}
	\mathcal{F}(\g_{12_\pm})=\{\varnothing, \{\g_{12_\pm}'\},\{\g_{12_\pm}\},\{\g_{12_\pm}',\g_{12_\pm}\}\}~.
\end{equation}

\subsection{Renormalization of the diagrams $\g_{11_\pm}$ and $\g_{12_\pm}$}
\quad\,
Let us start from the renormalization of $\g_{11_\pm}$ and $\g_{12_\pm}$. Despite presenting divergent subdiagrams, their renormalized integrands are simpler than those of $\g_{8_\pm}$ and $\g_{9_\pm}$. Taking into account the Zimmermann's forest formula (\ref{forest}), the renormalized
integrand for $\g_{11_\pm}$ reads:
\begin{equation}\label{fu11}
	R_{\g_{11_{\pm}}}(p,k_1,k_2,s)= \l_{11}e^2 S_{\g_{11_{\pm}}} \sum_{U\in \mathcal{F}_{\g_{11_{\pm}}}}\prod_{\l\in U}(-\t_\l S_\l )\tilde{I}^{\m\n}_\pm(k_1,k_2,p,s)~,
\end{equation}
where we have used (\ref{I11}) and (\ref{I-hat}). As already discussed, $\t_\l$ is related to the Taylor operator that acts on $p^\l$ and $s^\l$, being $p^\l$ the external momentum of the subdiagram whereas $s^\l$ is its Lowenstein-Zimmermann parameter.
Considering $\g_{11_\pm}'$, let us denote by $p^{\g_{11_\pm}'}$ its external momentum, $k^{\g_{11_\pm}'}$ its internal momentum and $s^{\g_{11_\pm}'}$ as its Lowenstein-Zimmermann parameter. 
In this case, it is necessary to know the results of applying the substitution operator $S_{\g_{11_\pm}'}$ in $\tilde{I}^{\m\n}_\pm(k_1,k_2,p,s)$, which means basically to determine $p^{\g_{11_\pm}'}$, $k^{\g_{11_\pm}'}$ and $s^{\g_{11_\pm}'}$ as functions of $p,~k_1,~k_2$ and $s$, allowing to change the latter variables by the variables of the subdiagram in the terms that correspond to it. It is possible to determine $p^{\g_{11_\pm}'}$ and $k^{\g_{11_\pm}'}$ following a similar approach as in Ref. \cite{Blaschke}.
The next figure shows how to relate the quantities of the diagram and its subdiagram: 
\vspace{0.4cm}
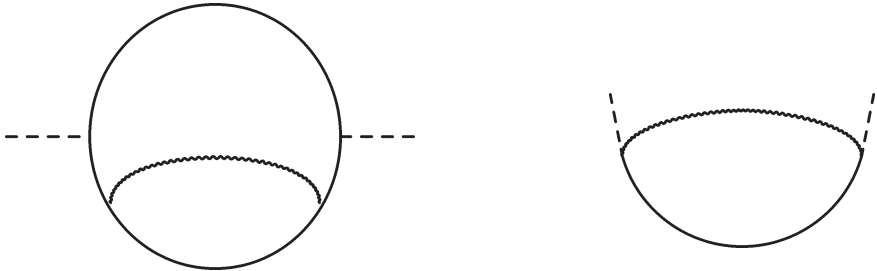
\begin{figure}[H]
	\center
	\begin{fmffile}{momentssub} 
		\begin{fmfgraph*}(55,35)
			\fmfset{wiggly_len}{18}
			\fmfipair{i,va,vb,vc,vd,o,c,ve,vf,ii}
			\fmfiequ{ii}{(0.05w,0.5h)}	
			\fmfiequ{i}{(0,.5h)}
			\fmfiv{label=$\gamma_{11_{\pm}}$,label.angle=80,label.dist=1.5cm}{i}
			\fmfiequ{va}{(.2w,.5h)}
			\fmfiequ{vb}{(.5w,.h)}
			\fmfiequ{vc}{(.8w,.5h)}
			\fmfiequ{vd}{(.5w,0)}
			\fmfiequ{o}{(w,.5h)}
			\fmfiequ{c}{(.5w,.5h)}
			\fmfiequ{ve}{c+(-.25w,-.25h)}
			\fmfiequ{vf}{c+(.25w,-.25h)}
			\fmfiv{label=$k_1$}{vb}
			\fmfiv{label=$p$,label.angle=90,label.dist=0.1cm}{ii}
			\fmfiv{label=$k_1-p$,label.angle=-65,label.dist=0.5cm}{vc}
			\fmfi{wiggly, lab=$k_2$}{ve{up} .. tension 2.3 .. {down}vf}
			\fmfiv{label=$k_1-k_2-p$}{vd}
			\fmfi{dashes}{i--va}
			\fmfi{dashes}{vc{right} .. {right}o}
			\fmfi{plain}{va{up} .. tension 1 .. {right}vb}
			\fmfi{plain}{vb{right} .. tension 1 .. {down}vc}
			\fmfi{plain}{vc{down} .. tension 1 .. {left}vd}
			\fmfi{plain}{vd{left} .. tension 1 .. {up}va}
		\end{fmfgraph*}
		\quad\quad\quad \quad \quad \quad
		\begin{fmfgraph*}(35,30)
			\fmfset{wiggly_len}{17}
			\fmfipair{i,o,va,vb,vi,vo}
			\fmfiequ{i}{(.05w,.5h)}
			\fmfiequ{o}{(0.95w,.5h)}
			\fmfiequ{va}{(0.5w,.7h)}
			\fmfiequ{vb}{(0.5w,.1h)}
			\fmfiequ{vi}{(0,.8h)}
			\fmfiequ{vo}{(w,.8h)}
			\fmfi{plain}{i.. tension 1 .. {right}vb}
			\fmfi{plain}{vb{right}.. tension 1 .. o}
			\fmfi{dashes}{i.. tension 1.5 .. vi}
			\fmfi{dashes, lab=$p^{\g_{11_\pm}'}$}{o.. tension 1.5 .. vo}
			\fmfi{wiggly}{i{up}.. tension 1.5 .. {right}va}
			\fmfi{wiggly}{va{right}.. tension 1.5 .. {down}o}
			\fmfiv{label=$\g_{11}'$,label.angle=90,label.dist=1.7cm}{i}
			\fmfiv{label=$k^{\g_{11_\pm}'}$}{va}
		\end{fmfgraph*}
	\end{fmffile}
	\vspace{0.4cm}
	\caption{Complete diagram $\g_{11_\pm}$ and its subdivergent diagram, with the momenta explicit.}
\end{figure}
\noindent
Comparing both, it is possible to see that
\begin{equation}\label{labelmoments}
	k^{\g_{11_\pm}'}=k_2~;~~~~~~~~~p^{\g_{11_\pm}'}=k_1-p~.
\end{equation}
Additionally, the parameters $s$ that are in the expression of $\g_{11_\pm}$
but result from the structure of $\g_{11_\pm}'$ must be replaced by $s^{\g_{11_\pm}'}$.

The application of the BPHZL procedure also requires the determination of the subtraction degrees $\d$ (UV) and $\r$ (IR), obeying the constraints (\ref{consistencyb}), (\ref{consistencyc}) and (\ref{vinc}). 
For the divergent subdiagram, it is possible to utilize those obtained for
the self-energy at 1-loop. Therefore, 
\begin{equation}\label{degrees11}
	\d(\g_{11_\pm}')=0~;~~~~~~~~~~~~\r(\g_{11_\pm}')=1~.
\end{equation}
For the complete diagram $\g_{11_\pm}$, there are the following constraints:
\begin{align}
	&\d\left(\g_{11_\pm}\right)=d\left(\g_{11_\pm}\right)+b\left(\g_{11_\pm}\right)~,\\[0.7em]
	&\d\left(\g_{11_\pm}\right) \geq d\left(\g_{11_\pm}/\g_{11_\pm}'\right)
	+\d\left(\g_{11_\pm}'\right)~,\\[0.7em]
	&\r\left(\g_{11_\pm}\right)=r\left(\g_{11_\pm}\right)-c\left(\g_{11_\pm}\right)~,\\[0.7em]
	&\r\left(\g_{11_\pm}\right)\leq r\left(\g_{11_\pm}/\g_{11_\pm}'\right)+
	\r\left(\g_{11_\pm}'\right)~,\\[0.7em]
	&\r\left(\g_{11_\pm}\right)\leq \d\left(\g_{11_\pm}\right)+1~.
\end{align}
The reduced diagram, when $\g_{11_\pm}'$ is contracted to a point, is shown in the following figure:
\begin{figure}[H]\label{reduced}
	\center
	\begin{fmffile}{reduced} 
		\begin{fmfgraph*}(55,35)
			\fmfset{wiggly_len}{18}
			\fmfipair{i,va,vb,vc,vd,o,c,ve,vf,ii}
			\fmfiequ{ii}{(0.05w,0.5h)}	
			\fmfiequ{i}{(0,.5h)}
			\fmfiv{label=$\gamma_{11_{\pm}}/\gamma_{11_{\pm}}'$,label.angle=80,label.dist=1.5cm}{i}
			\fmfiequ{va}{(.2w,.5h)}
			\fmfiequ{vb}{(.5w,.h)}
			\fmfiequ{vc}{(.8w,.5h)}
			\fmfiequ{vd}{(.5w,0)}
			\fmfiequ{o}{(w,.5h)}
			\fmfiequ{c}{(.5w,.5h)}
			\fmfiequ{ve}{c+(-.25w,-.25h)}
			\fmfiequ{vf}{c+(.25w,-.25h)}
			\fmfiv{d.sh=circle,d.f=1,d.si=5pt}{vd}
			\fmfi{dashes}{i--va}
			\fmfi{dashes}{vc{right} .. {right}o}
			\fmfi{plain}{va{up} .. tension 1 .. {right}vb}
			\fmfi{plain}{vb{right} .. tension 1 .. {down}vc}
			\fmfi{plain}{vc{down} .. tension 1 .. {left}vd}
			\fmfi{plain}{vd{left} .. tension 1 .. {up}va}
		\end{fmfgraph*}
	\end{fmffile}
	\vspace{0.4cm}
	\caption{Reduced subdiagram.}
\end{figure}
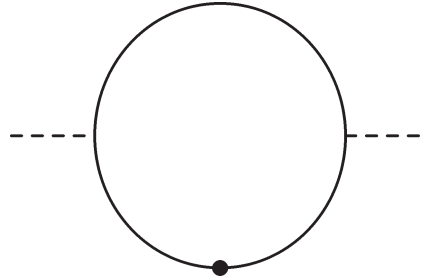
To derive $d(\g_{11_\pm}/\g_{11_\pm}')$ and $r(\g_{11_\pm}/\g_{11_\pm}')$, it is still
possible to use the power-counting formula (\ref{power_counting}), bearing in mind
that there is the addition of one fermionic propagator, due to the contraction
of the divergent subdiagram, which reduces one unit of $d$ and $r$. Thus, it is possible to see that
\begin{equation}
	d(\g_{11_\pm}/\g_{11_\pm}')=0~,~~~~~~~~r(\g_{11_\pm}/\g_{11_\pm}')=0
\end{equation}
So, the constraints become
\begin{equation}
	\begin{split}
		&\d\left(\g_{11_\pm}\right)=d\left(\g_{11_\pm}\right)+b\left(\g_{11_\pm}\right)~,\\[0.7em]
		&\d\left(\g_{11_\pm}\right) \geq
		+\d\left(\g_{11_\pm}'\right)~,\\[0.7em]
		&\r\left(\g_{11_\pm}\right)=r\left(\g_{11_\pm}\right)-c\left(\g_{11_\pm}\right)~,\\[0.7em]
		&\r\left(\g_{11_\pm}\right)\leq 
		\r\left(\g_{11_\pm}'\right)~,\\[0.7em]
		&\r\left(\g_{11_\pm}\right)\leq \d\left(\g_{11_\pm}\right)+1~.
	\end{split}
\end{equation}
So, taking $c(\g_{11_\pm})=b(\g_{11_\pm})=0$, it follows that
\begin{equation}
	\d(\g_{11_\pm})=0~,~~~~~~~~\r(\g_{11_\pm})=1~.
\end{equation}

The next step is to determine the Taylor operators $\t_{\g_{11_\pm}}$ and $\t_{\g_{11_\pm}'}$. Since the subtraction degrees are the same for both the
complete diagram and the subdiagram, determining $\t_{\g_{11_\pm}}$ automatically determines $\t_{\g_{11_\pm}'}$. For
$\g_{11_\pm}$, the Taylor operator reads
\begin{align}
	\nonumber
	1-\t_{\g_{11}}=&\left(1-t^{\r(\g_{11_\pm})-1}_{p,(s-1)}\right)\left(1-t^{\d(\g_{11_\pm})}_{p,s}\right) \\
	\nonumber
	=& 1-t^0_{p,(s-1)}-t^0_{p,s}+t^0_{p,(s-1)}t^0_{p,s} \\
	=& 1-t^0_{p,(s-1)}~.
\end{align}
The aforementioned expression was obtained using the fact that  $t^0_{p,s}$ annihilates $p~\textrm{and }s$, such that $t^0_{p,(s-1)}t^0_{p,s}=t^0_{p,s}$, because $t^0_{p,(s-1)}$ acts on a expression that has neither $p$ nor $s$, effectively becoming the identity operator.
It can be concluded that:
\begin{equation}
	\t_{\g_{11}}=t^0_{p,(s-1)}~,~~~~~~~~\t_{\g_{11}'}=t^0_{p^{\g_{11}'},(s^{\g_{11}'}-1)}~.
\end{equation}

Now we are in position of determining the renormalized integrand for $\g_{11}$. Expanding the expression (\ref{fu11}), we get
\begin{align}\label{ren}
	\frac{R^{\g_{11}}_ {\m\n}(p,k_1,k_2,s)}{\l_{11} e^2}=[1-\t_{\g_{11}}
	-S_{\g_{11}}\t_{\g_{11}'}S_{\g_{11}'}-
	\t_{\g_{11}}S_{\g_{11}}\t_{\g_{11}'}S_{\g_{11}'}]\tilde{I}_{\m\n}(p,k_1,k_2,s)~.
\end{align}
It remains to determine the terms of $\widetilde{I}^{\m\n}(p,k_1,k_2,s)$ which
correspond to the subdiagram $\g_{11}'$. Observing the expression in (\ref{I-hat}),
it is possible to see that it is situated between $\g_\a$ and $\g_\b$. Using the trace properties, together with (\ref{labelmoments}),
it is possible to obtain the following equality:
\begin{equation}
	S_{\g_{11}'}\widetilde{I}^{\m\n}(p,k_1,k_2,s)=-{\rm Tr}\{{I}^{\m\n}_{\g_{11}/\g_{11}'}(p,k_1,s)\,{I}^{\m\n}_{\g_{11}'}(p^{\g_{11}},k^{\g_{11}},s^ {\g_{11}})\}~,
\end{equation}
where
\begin{align}
	\nonumber
	{I}^{\m\n}_{\g_{11}/\g_{11}'}(p,k_1,s)=	&
	\left[ i \frac{(\Sl k_1- \Sl p)\mp m(s-1)}{(k_1-p)^2-m^2(s-1)^2} \right]\g^\m \left[ i \frac{\Sl k_1\mp m(s-1)}{k_1^2-m^2(s-1)^2} 
	\right] \times\\
	& \times
	\gamma^{\n}
	\left[ i \frac{(\Sl k_1- \Sl p)\mp m(s-1)}{(k_1-p)^2-m^2(s-1)^2} \right] ~,
\end{align}
while
\begin{align}
	\nonumber
	{I}^{\m\n}_{\g_{11}'}(p^{\g_{11}},k^{\g_{11}},s^ {\g_{11}})=&
	\g_\a\left[-i\frac{1}{(k^{\g_{11}'})^2-\mu^2}
	\left(\eta^{\alpha \beta}-\frac{(k^{\g_{11}'})^{\alpha}(k^{\g_{11}'})^{\beta}}
	{(k^{\g_{11}'})^2}\right)\right] \times \\
	&\times \left[ i \frac{ (\Sl p^{\g_{11}'})- (\Sl k^{\g_{11}'})\mp m(s-1)}{((p^{\g_{11}'})- (k^{\g_{11}'}))^2-m^2(s-1)^2} \right] \gamma_{\b}~.
\end{align}
Now it is possible to expand the non-trivial terms of (\ref{ren})  and take the limit $s\rightarrow 1$ to recover the massless regime, so that we are left with:
\begin{equation}\label{int01}
	\tau_{\g_{11}} \widetilde{I}_{\g_{11}} = i \Tr[\gamma_\theta \gamma_\mu \gamma_\omega \gamma_\nu \gamma_\xi \gamma_\alpha \gamma_\chi \gamma_\beta] \frac{k_1^\theta k_1^\omega k_1^\xi(k_1^\chi - k_2^\chi)}{k_1^2k_1^2k_1^2(k_1 - k_2)^2}\frac{1}{k_2^2 - \mu^2}\left(\eta^{\alpha\beta} - \frac{k_2^\alpha k_2^\beta}{k_2^2}\right)~,
\end{equation}

\begin{equation}\label{int03}
	\begin{split}
		S_{\g_{11}} (\tau_{\g_{11}'}S_{\g_{11}'} {\wt I}_{\g_{11}})|_{s=1} &= i \Tr[\gamma_\theta \gamma_\mu \gamma_\omega \gamma_\nu \gamma_\xi \gamma_\alpha \gamma_\chi \gamma_\beta]\frac{(k_1^\theta - p^\theta)(k_1^\xi - p^\xi)}{((k_1 -p)^2)^2}
		\frac{k_1^\omega}{k_1^2}\frac{k_2^\chi}{k_2^2}\\&\frac{1}{k_2^2 - \mu^2}\left(\eta^{\alpha\beta} - \frac{k_2^\alpha k_2^\beta}{k_2^2}\right)~.
	\end{split}
\end{equation}

\begin{equation}\label{int02}
	\tau_{\g_{11}} S_{\g_{11}} ( \tau_{\g_{11}'}S_{\g_{11}'} {\wt I}_{\g_{11}}) = i \Tr[\gamma_\theta \gamma_\mu \gamma_\omega \gamma_\nu \gamma_\xi \gamma_\alpha \gamma_\chi \gamma_\beta]\frac{k_1^\theta k_1^\omega k_1^\xi}{k_1^2k_1^2k_1^2}\frac{k_2^\chi}{k_2^2}\;\frac{1}{k_2^2 - \mu^2}\left(\eta^{\alpha\beta} - \frac{k_2^\alpha k_2^\beta}{k_2^2}\right)~.
\end{equation}

\begin{equation}
	\begin{split}\label{int04}
		I_\pm^{\mu\nu}(p,k_1,k_2,s=1) &= i\Tr[\gamma_\theta \gamma_\mu \gamma_\omega \gamma_\nu \gamma_\xi \gamma_\alpha \gamma_\chi \gamma_\beta]\frac{( k_1^\theta - p^\theta)}{(k_1 - p)^2}\frac{k_1^\omega}{k_1^2 }\frac{( k_1^\xi - p^\xi)}{(k_1 - p)^2}\frac{( k_1^\chi -  k_2^\chi - p^\chi)}{(k_1 - k_2 - p)^2}\times\\
		&\times\frac{1}{k_2^2 - \mu^2}\left(\eta^{\alpha\beta} - \frac{k_2^\alpha k_2^\beta}{k_2^2}\right)~,
	\end{split}
\end{equation}

The renormalization of $\g_{11_{\pm}}$ by BPHZL now is summarized considering the expressions (\ref{int04}), (\ref{int01}), (\ref{int02}) and (\ref{int03}). Several results which are presented in Appendix \ref{ApendiceB} will be used to solve the integrals, such as the Feynman parametrization and the $J_r$ integrals. It is worth emphasizing that these expressions do not break parity, because  there are no traces of an odd number of $\g$ matrices, which agrees with the results of Ref. \cite{parity}. 

It must be recalled that the integrands are integrated with respect to the momenta $k_1$ and $k_2$. Regarding the subtraction terms originated from the divergent subdiagram, it is possible to see that (\ref{int02}) has a power three in the momentum $k_1$ in the numerator. After the Feynman parametrization and considering the integration over the internal momenta, it can be written as
\begin{equation}
	\int \frac{d^3 k_2 }{(2\pi)^3} \frac{d^3 k_1 }{(2\pi)^3} \frac{k_1^\theta k_1^\omega k_1^\xi f(k_2)}{(k_1^2 -c)^\a}~,
\end{equation}
where $f(k_2)$ encompasses all the other terms, such as the integration in the Feynman parameters and $c$ is independent of the momentum $k_1$. Therefore, the integration with respect to $k_1$ vanishes, because there is an odd number of momentum $k_1$ in the numerator and there is no term of the form $k_1\cdot p$ in the denominator. By a similar argument, considering the integration over $k_2$, it follows that the contribution of (\ref{int03}) vanishes. Thus, the subtraction terms due to the divergent subdiagram vanish. It remains the evaluation of expressions (\ref{int01}) and (\ref{int04}), which effectively contribute to the renormalized integrand.

In this context, the renormalized diagram becomes
\begin{equation}\label{Renn}
	\begin{split}
		\frac{\g_{11}^{(R)}}{\l_{11}e^2}&= 
		\int\frac{d^3k_1}{(2\pi)^3} \frac{d^3k_2}{(2\pi)^3} \left(\widetilde{I}_{\m\n}(p,k_1,k_2,1)-\t_{\g_{11}} \widetilde{I}_{\m\n}(p,k_1,k_2,1)\right)
		\\
		&=i\Tr[\gamma_\theta \gamma_\mu \gamma_\omega \gamma_\nu \gamma_\xi \gamma_\alpha \gamma_\chi \gamma_\beta]\int \frac{d^3k_1}{(2\pi)^3} \frac{d^3k_2}{(2\pi)^3}
		\times \\
		&\times\frac{k_1^\theta k_1^\xi (k_1^\chi-k_2^\chi)(\eta^{\a\b}k_2^2-k_2^\a k_2^\b)(2k_1^\o k_1^\s p_\s-k_1^\o p^2-p^\o k_1^2)}{(k_1^2)^3(k_1-p)^2 k_2^2 (k_1-k_2)^2(k_2^2-\m^2)}~.
	\end{split}
\end{equation}
As can be checked, the previous integral is convergent, since it behaves as $1/k$ at large momenta.

The following Feynman parametrization (see Equation (\ref{pfeynmangeral})) is necessary:
\begin{equation}
	\begin{split}
		&\frac{1}{(k_1^2)^3(k_1-p)^2 k_2^2 (k_1-k_2)^2(k_2^2-\m^2)} = \frac{\G(7)}{\G(3)}
		\int dx\, dy\, dv\, dw\, dz\, \times\\
		&\times\frac{\d(1-x-y-v-w-z) x^2}{(xk_1^2+y((k_1-p)^2)+v k_2^2+w (k_1-k_2)^2+z (k_2^2-\m^2))^7}.
	\end{split}
\end{equation}
Here, the Feynman parameters are $x~,y,~v,~w$ and $z$ and they range from 0 to 1. Using the previous result,
along with (\ref{Renn}), and integrating over the parameter $v$, we obtain the following expression for the renormalized diagram:
\begin{equation}
	\frac{\g_{11}^{(R)}}{\l_{11}e^2}=
	\int d\Theta \frac{d^3 k_1}{(2\pi)^3} \frac{d^3 k_2}{(2\pi)^3}\frac{k_1^\theta k_1^\xi (k_1^\chi-k_2^\chi)(\eta^{\a\b}k_2^2-k_2^\a k_2^\b)(2k_1^\o k_1^\s p_\s-k_1^\o p^2-p^\o k_1^2)}{(k_2^2 + 2 p'\cdot k_2 -c)^7}~.
\end{equation}
where
\begin{equation}
	d\Theta = dx\, dy\, dw\, dz \frac{x^2}{(1-x-y)^7} i\Tr[\gamma_\theta \gamma_\mu \gamma_\omega \gamma_\nu \gamma_\xi \gamma_\alpha \gamma_\chi \gamma_\beta]~\frac{\G(7)}{\G(3)}~,
\end{equation}
whereas
\begin{equation}
	p'  = -\frac{w}{1-x-y}k_1 , \qquad c' = \frac{1}{1-x-y}(\mu^2 z - (w+x+y)k_1^2 - 2y k_1\cdot p - y p^2)~.
\end{equation}
It will be convenient for later use to define
\begin{equation}
	a= -\frac{w}{1-x-y}~.
\end{equation}

From the results of the Appendix \ref{ApendiceB}, integrating first over $k_2$,
it follows that
\begin{equation}\label{fin}
	\frac{\g_{11}^{(R)}}{\l_{11}e^2}= \int d\Theta \frac{d^3 k_1}{(2\pi)^3}
	\frac{i(-1)^7 \pi^{3/2}}{\Gamma(7)(2\pi)^3} k_1^\theta k_1^\xi \eta_{\rho\sigma} ( - k_1^\rho k_1^\sigma p^\omega +2k_1^\rho k_1^\omega p^\sigma  - k_1^\omega p^\rho p^\sigma )[I^{\alpha\beta\chi}_{k_2}(J_2) + I^{\alpha\beta\chi}_{k_2}(J_3)]
	~,
\end{equation}
where
\begin{equation}
	I^{\alpha\beta\chi}_{k_2}(J_2) = \left[(c'+{p'}^2)^{-\frac{11}{2}}\Gamma\left(\frac{11}{2}\right)({p'}^2 \eta^{\alpha\beta}-{p'}^\alpha {p'}^\beta) - (c'+{p'}^2)^{-\frac{9}{2}}\Gamma\left(\frac{9}{2}\right)\eta^{\alpha\beta} \right]k_1^\chi~,
\end{equation}
and
\begin{equation}
	\begin{split}
		I^{\alpha\beta\chi}_{k_2}(J_3) &= -\left[-(c'+{p'}^2)^{-\frac{11}{2}}\Gamma\left(\frac{11}{2}\right){p'}^\chi ({p'}^2 \eta^{\alpha\beta}-{p'}^\alpha {p'}^\beta) +\right. \\
		&+\left. \frac{1}{2}(c'+{p'}^2)^{-\frac{9}{2}}\Gamma\left(\frac{9}{2}\right)(4\eta^{\alpha\beta}{p'}^\chi - \eta^{\chi\alpha}{p'}^\beta - \eta^{\beta\chi}{p'}^\alpha)  \right]~.
	\end{split}
\end{equation}
It is worth noting that $\eta_{\r\s}$ appeared in (\ref{fin}) to write all the momenta
with contravariant components. This will also be used in the following, 
making ${p'}^2=\eta_{\kappa \gamma}{p'}^\kappa{p'}^\g $.

For simplification, the change $k_1 \rightarrow k$ is made. Substituting $c'$ and $p'$, we have
\begin{equation}\label{finalintegral}
	\begin{split}
		c' + {p'}^2  
		= u\left(k^2 + 2k\cdot p'' - c''\right)~,
	\end{split}
\end{equation}
with $u$, $p''$ and $c''$ given by:
\begin{equation}
	\begin{split}
		&u =  \left(\frac{w^2 - (1-x-y)(w+x+y)}{(1-x-y)^2}\right) \qquad p'' = - \frac{2y(1-x-y)}{w^2 - (1-x-y)(w+x+y)} p\\
		&\qquad \qquad \qquad \qquad \qquad c'' = \frac{(1-x-y)(y p^2-z \mu^2)}{w^2 - (1-x-y)(w+x+y)}
	\end{split}
\end{equation}
The form of (\ref{finalintegral}) allows to integrate over $k$, using the $J_r$ integrals once more. Finally, the renormalized diagram is given by the expression
\begin{equation}\label{rg11}
	\begin{split}
		\frac{\g_{11}^{(R)}}{\l_{11}e^2}=
		&\int d\Theta' \eta_{\rho\sigma}\left\{\Gamma\left(\frac{11}{2}\right)\frac{a^2(1+a)}{u^{\frac{11}{2}}}\left[-\left(\eta^{\alpha\beta}\eta_{\kappa\gamma} J_7^{\theta\xi\rho\sigma\kappa\gamma\chi}\right.\right.\right.+ \\
		&- \left. J_7^{\theta\xi\rho\sigma\alpha\beta\chi}\right) p^{\omega} +2\left(\eta^{\alpha\beta}\eta_{\kappa\gamma} J_7^{\theta\xi\rho\omega\kappa\gamma\chi}\right. +\\
		&-\left.\left. J_7^{\theta\xi\rho\omega\alpha\beta\chi} \right)p^{\sigma} -\left( \eta^{\alpha\beta}\eta_{\kappa\gamma} J_6^{\theta\xi\omega\kappa\gamma\chi} -  J_6^{\theta\xi\omega\alpha\beta\chi} \right)p^{\rho}p^\sigma\right]+\\
		&- \Gamma\left(\frac{9}{2}\right)\frac{1}{2u^{\frac{9}{2}}}\left[ -\left(2(1+2a)\eta^{\alpha\beta} J_5^{\theta\xi\rho\sigma\chi} -a\eta^{\chi\alpha} J_5^{\theta\xi\rho\sigma\beta} \right.\right.+
		\\&- \left. a\eta^{\beta\chi} J_5^{\theta\xi\rho\sigma\alpha}\right)p^\omega +2\left(2(1+2a)\eta^{\alpha\beta} J_5^{\theta\xi\rho\omega\chi} -a\eta^{\chi\alpha} J_5^{\theta\xi\rho\omega\beta} \right.+
		\\&- \left. a\eta^{\beta\chi} J_5^{\theta\xi\rho\omega\alpha}\right)p^\sigma - \left(2(1+2a)\eta^{\alpha\beta} J_4^{\theta\xi\omega\chi} -a\eta^{\chi\alpha} J_4^{\theta\xi\omega\beta} \right.+
		\\&- \left. \left. \left. a\eta^{\beta\chi} J_4^{\theta\xi\omega\alpha}\right)p^\rho p^\sigma \right] \right\}
	\end{split}
\end{equation}
with $J_7$ and $J_6$ being functions of $p'', c'', \alpha=11/2$, while $J_5$ and $J_4$ are functions of $p'', c'', \alpha=9/2$. Furthermore, $d \Theta'$ is $d\Theta$
with the constant factors absorbed, encompassing all the integration elements of the Feynman
parameters $x,~y,~w$ and $z$.

We now shall prove that all the terms resulting from the renormalized diagram $\g_{11_\pm}$
are non-local. From (\ref{generalJ}), the terms that contain a $J_7$ integral are given by ($\lfloor \cdot \rfloor$ denotes the floor function):
\begin{equation}
	\begin{split}
		J_7^{\m_1 ... \m_r}(11/2,p'',c'')&=i(-1)^{25/2} \frac{\pi^{3/2}}{\G(\a)(2\pi)^3}\sum_{j=0}^{\lfloor \frac{7}{2} \rfloor}\left(-\frac{1}{2}\right)^j \G(11/2-3/2-j) \times\\
		&\times (c''+{p''}^2)^{3/2-11/2+j}\bigg(\O(p)_{7-2j}^{\m_1 ... \m_{7-2j}} \zeta_{2j}^{\m_{7-2j+1}... \m_7}+\textrm{dist. perm.}\bigg)~~.
	\end{split}
\end{equation}
The highest value of $j$ is $\lfloor 7/2 \rfloor=3$. Therefore, the term
$(c''+{p''}^2)$ has the maximum power given by $-1$, which results in a denominator
containing terms with $p^2$, where $p$ is the external momentum, which yields non-local
terms in configuration space. By a similar argument, it is also possible to check that
the all terms containing $J_6$, $J_5$ and $J_4$ in (\ref{rg11}) yield non-local terms. Thus, the renormalized diagram $\g_{11_{\pm}}$ does not contribute to the renormalization of any parameters of the model. 

The BPHZL renormalization of $\g_{12_{\pm}}$ is entirely analogous, and allows us to conclude that this diagram will also yield only non-local terms.

\subsection{Renormalization of the diagrams $\g_{8_\pm}$ and $\g_{9_\pm}$}
\quad \,
Although the expressions of the forests corresponding to $\gamma_{8_\pm}$ and $\gamma_{9_\pm}$ are simpler than those for $\gamma_{10_\pm}$ and $\gamma_{11_\pm}$, their renormalization procedures are more laborious due to the greater number of terms that are generated. It is worth noting that the expressions of $\gamma_{8_\pm}$ and $\gamma_{9_\pm}$ are the same, except for a constant factor. Hence, it is sufficient to renormalize only one of them. We choose to perform the renormalization of $\gamma_{8_\pm}$.

Using the Zimmermann's forest formula and the same subtraction degrees in (\ref{degrees11}),
we obtain
\begin{align}
\nonumber
    R^{\m\n}_{\g_{8_{\pm}}}(p,k_1,k_2,s) =& \l_8 e^2 S_{\g_{8_{\pm}}} \sum_{U\in\mathcal{F}_{\g_{8_{\pm}}}}\prod_{\l\in U}(-\t_\l S_\l)\widehat{I}^{\m\n}_\pm(k_1,k_2,p,s) \\
    =& \l_8 e^2 (1-t^0_{p,(s-1)}) \widehat{I}^{\m\n}_\pm(k_1,k_2,p,s)~.
\end{align}
where $\widehat{I}_{\m\n}$ is given by (\ref{I-tilde}). After performing the calculations and taking $s \rightarrow 1$ to recover
the massless case, the previous expression reads
\begin{equation}\label{Rg8}
\begin{split}
    &R^{\m\n}_{\g_{8_{\pm}}}(p,k_1,k_2,1)= i \l_8 e^2 \textrm{Tr}
     [\g^\m \g_\theta \g_\a \g_\chi \g^\n \g_\xi \g_\b \g_\o] \times\\[2ex]
   &\quad \times\frac{k_1^\theta k_2^\chi[(k_1-k_2)^2\eta^{\a\b}-(k_1^\a - k_2^\a)(k_1^\b - k_2^\b)]}{k_1^2((k_1-k_2)^2-\m^2)(k_1-k_2)^2 k_2^2} \times
   \\[2ex]
   &\quad\times
   \frac{-k_1^2 k_2^2(k_1^\o p^\xi +k_2^\xi p^\o - p^\o p^\xi)-k_1^\o k_2^\xi\left(k_1^2(-2k_2 \cdot p + p^2)+(-2k_1 \cdot p +p^2)(k_2-p)^2\right)}{k_1^2 k_2^2(k_1-p)^2(k_2-p)^2}
\end{split}
\end{equation}
Considering the highest power of $k$ in the numerator and in the denominator of (\ref{Rg8}), it is possible
to check that it is a convergent integral,
because $\displaystyle d^3k_1 d^3k_2 R^{\m\n}_{\g_{8_\pm}}\sim 1/k$. 

We integrate the renormalized integrand first with respect to the momentum $k_1$, using the results in Appendix \ref{ApendiceB}. Therefore, the denominator must take the form $(k_1^2+2 k\cdot p' - c')^\alpha$, for some ${p'}^\m$ and $c'$ that depend on $k_2^\m, p^\m, \m$ and numerical factors. Using the generalized Feynman parametrization (\ref{pfeynmangeral}), we obtain
\begin{equation}
  \begin{split}
  		&\frac{1}{(k_1)^2(k_2)^2[(k_1-k_2)^2-\m^2](k_1-k_2)^2(k_1-p)^2(k_2-p)^2} = \\[2ex]
  		&= \int  \frac{dx_1 dx_2 dx_3 dx_4 dx_5 dx_6 ~~ \G(8)~\d(1-x_1 ... -x_6)~ x_1 x_2}{\big[k_1^2 x_1+ k_2^2 x_2 +[(k_1-k_2)^2-\m^2]x_3+(k_1-k_2)^2x_4+(k_1-p)^2x_5+(k_2-p)^2x_6\big]^8} \\[2ex]
  		&=\G(8)\int dx_1 dx_2 dx_4 dx_5 dx_6 \frac{x_1 x_2}{(1-x_2)^8}~ \frac{1}{(k_1^2+2k_1\cdot p' -c')^8}~,
  \end{split}	
\end{equation}
where the Feynman parameters are denoted as $x_1,~...,x_6$, and we have chosen to eliminate the parameter $x_3$ by integrating the distribution $\d$. Additionally, 
we have $\a=8$ and the quantities ${p'}^\m$ and $c'$ are
\begin{equation}\label{p,k}
	\begin{split}
		&{p'}^\m = \frac{\left[(x_1+x_2+x_5+x_6-1)k_2^\m-x_5 p^\m\right]}{1-x_2} \\
		&c' = \frac{(x_1+x_5-1)~k_2^2+2x_6~ k_2\cdot p+(1-x_1-x_2-x_4-x_5-x_6)\m^2-(x_5+x_6)p^2}{1-x_2}
	\end{split}
\end{equation}

Next, we turn our attention to the numerator of (\ref{Rg8}). We will expand it in terms of powers of the momentum $k_1$. Each of these terms, after integrating with respect to
this internal momentum, will correspond to a term containing a $J_r(8,p',c')$, being $r$ the power
of $k_1$. The number of terms after the expansion is large, so it is convenient to
introduce a new notation to make the expressions more readable. We define
\begin{equation}
	f_n^{\m_1 ... \m_n}= k_1^{\m_1}...k_1^{\m_n}~~;~~~~~~~~~ g_n^{\m_1 ... \m_n}= k_2^{\m_1}...k_2^{\m_n}~~.
\end{equation}
Additionally, if the subscript is greater than the number of indices, there is an implicit
scalar product. For instance,
\begin{equation}
	f_3^{\m} = k_1^{\m}  k_1^2~.
\end{equation}
Furthermore, the following properties hold:
\begin{equation}
	f_n^{\m_1... \m_n} f_m^{\m_n... \m_{m+n}} = f_{m+n}^{\m_1... \m_{m+n}}~~;~~~~~~~~
	g_n^{\m_1... \m_n} g_m^{\m_n... \m_{m+n}} = g_{m+n}^{\m_1... \m_{m+n}}
\end{equation}
With all these considerations, the numerator in (\ref{Rg8}) can be expanded as
$\displaystyle{\sum_{l=0}^6 X_l}$, where
\begin{equation}\label{X}
	\begin{split}
		X_0&=\bigg[-f_6^{\theta \o}g_3^\chi p^\xi-f_5^{\theta}g_4^{\xi \chi} p^\o+f_5^\theta g_3^\chi p^\o p^\xi+2f_6^{\theta \o}g_3^{\chi \xi \d} p_\d -f_6^{\theta \o}g_2^{\chi\xi}p^2+\\
		&~~~2 f_5^{\theta \o \d}\left(g_4^{\chi\xi}-2g_3^{\chi\xi\d}p_\d+g_2^{\chi\xi}p^2\right)p_\d
		-f_4^{\theta\o}\left(g_4^{\chi\xi}-2g_3^{\chi\xi\d}p_\d+g_2^{\chi\xi}p^2\right) p^2\bigg]\eta^{\a\b}~; \\[4ex]
		X_1&=-\frac{f_1^{\d'}g_{1\d'}}{f_2} X_0~;~~~~
		X_2 =\frac{g_2}{f_2} X_0~,~~~~X_3= -\frac{f_2^{\a\b}}{f_2 \eta^{\a\b}} X_0~;
		~~~~X_4= \frac{f_1^{\a}g_1^{\b}}{f_2 \eta^{\a\b}} X_0~; \\[2ex]
		X_5&= \frac{f_1^{\b}g_1^{\a}}{f_2 \eta^{\a\b}} X_0~;~~~~X_6=-\frac{g_2^{\a\b}}{f_2\eta^{\a\b}} X_0~~.
	\end{split}
\end{equation}
We bypass the tensor structure of the $X_l$ for simplicity. Thus, the expression for the renormalized integrand (\ref{Rg8}) can be rewritten as
\begin{equation}
	R^{\m\n}_{\g_{8_\pm}}=\sum_{l=0}^6\int d\O \frac{X_l}{\left(k_1^2+2 k_1\cdot p' - c'\right)^8}~~,
\end{equation}
where
\begin{equation}
	d\O= \l_8 e^2 \textrm{Tr}\left(\g^\m\g_\theta \g_\a \g_\chi \g^\n \g_\xi \g_\beta \g_\o\right)
	\G(8) dx_1 dx_2 dx_4 dx_5 dx_6 \frac{x_1 x_2}{(1-x_2)^8}~.
\end{equation}
For $\g_{9_\pm}$, $d\O$ is the same, apart from  $\l_8 e^2$, which must be changed to $\l_9 g^2$.

Let us compute the contributions of $\g_{8_\pm}$ after
the BPHZL renormalization procedure. For this purpose, it is convenient to introduce
another notation:
\begin{align}
	\label{prodproperty}
	&\{p_i k_{n-i}\}^{\a_1 ... \a_n} = p^{\a_1}\cdots p^{\a_i} k^{\a_{i+1}}...k^{\a_n} + \textrm{dist. perm}~,\\[3ex]
&	(Ap^{\a_1} + B k^{\a_1})\cdots (A p^{\a_n} + B k^{\a_n})=\sum_{i=0}^n A^i B^{n-i} \{p_i k_{n-i}\}^{\a_1 ... \a_n}~,
\end{align}
where $A,~B$ are numbers, and by "dist. perm." we mean all the distinct permutations,
which are obtained by exchanging the tensor indices between $k$ and $p$. We refer to $\{\cdot\}^\cdot$ as an \textbf{expansion operator}. Furthermore,
\begin{equation}
	\begin{split}
	\{p_i k_{n-i}\}^{\a_1 ... \a_n\bar{\beta}_k} =& k^\beta(p^{\a_1}\cdots p^{\a_i} k^{\a_{i+1}}...k^{\a_n} + \textrm{dist. perm})\\
	& k^\beta \{p_i k_{n-i-1}\}^{\a_1 ... \a_n}~,
	\end{split}
\end{equation}
that is, an index with an upper bar is not exchanged.
In addition, if the sum of the subscripts is greater than the number of indices,
there is a contraction omitted in the quantity whose subscript is responsible for the asymmetry. For instance, $\{p_1 k_3\}^{\a \b}= p^\a k^\beta k^2 + \textrm{dist. perm}$. The generalization of the expansion operator for more than
two quantities is straightforward. Furthermore, the expansion operator  can be
used for the $J_r$ integrals discussed in Appendix \ref{ApendiceB}.
So, considering the previous definitions, we must calculate
\begin{equation}
	\int \frac{d^3k_1}{(2\pi)^3} \frac{d^3k_2}{(2\pi)^3} R^{\m\n}_{\g^{8_\pm}}
	= \sum_{l=0}^6\int d\O\int \frac{d^3k_2}{(2\pi)^3} \frac{d^3k_1}{(2\pi)^3}
	\frac{X_l}{\left(k_1^2+2 k_1\cdot p' - c'\right)^8}~~,
\end{equation}
After integrating in the momentum
$k_1$, using the definition of the $J_r$ integrals in Appendix \ref{ApendiceB}, the contribution of $X_0$ is
\begin{equation}\label{X'0}
\begin{split}
X'_0=\int \frac{d^3 k_1}{(2\pi)^3} \frac{X_0}{(k_1^2 + 2k_1 \cdot p' -c')^8} = \bigg[-J_6^{\theta \o}g_3^\chi p^\xi-J_5^{\theta}g_4^{\xi \chi} p^\o+J_5^\theta g_3^\chi p^\o p^\xi+2J_6^{\theta \o}g_3^{\chi \xi \d} p_\d \\ -J_6^{\theta \o}g_2^{\chi\xi}p^2+
2 J_5^{\theta \o \d}\left(g_4^{\chi\xi}-2g_3^{\chi\xi\d'}p_\d'+g_2^{\chi\xi}p^2\right)p_\d
-J_4^{\theta\o}\left(g_4^{\chi\xi}-2g_3^{\chi\xi\d}p_\d+g_2^{\chi\xi}p^2\right) p^2\bigg]\eta^{\a\b}~,
\end{split}
\end{equation}
where all the $J_r$ are functions of $\a=8$, $p'$ and $c'$.
Since the integral in $k_1$ have been performed, it remains only the internal momentum
$k_2$, which can be relabeled as $k$. Furthermore, 
it is possible to write $X'_0$ as
\begin{equation}
	X'_0=\sum_{l'=0}^{10} X'_{0l'}~.
\end{equation}
Each $X'_{0l'}$ corresponds to the $l'$-th term in the sum of (\ref{X'0}).
The first term, using (\ref{generalJ}), is
\begin{equation}
	\begin{split}\label{X'00}
	X'_{00}&= -J_6^{\theta \o}g_3^{\chi}p^\xi = -\eta_{\phi \phi'}\eta_{\psi \psi'}J_6^{\theta \o \phi \phi' \psi \psi'}(8,p',c')k^{\chi} k^2 p^\xi \eta^{\a\b} \\
	&=-\eta_{\phi \phi'}\eta_{\psi \psi'} k^{\chi} k^2 p^\xi C(8) \sum_{j=0}^3 \left(-\frac{1}{2}\right)^j \G(13/2-j)(c'+{p'}^2)^{-13/2+j}\times \\
	&~~~~\times \bigg(\O(p')_{6-2j}^{\m_1...\m_{6-2j}}\zeta_{2j}^{\m_{6-2j+1}... \m_6}+\textrm{dist. perm.}\bigg)\eta^{\a\b} ~~.
	\end{split}
\end{equation}
where $C(8)=i \dfrac{\pi^{3/2}}{\G(8)(2\pi)^3}$ and $(\m_1, ..., \m_6)= (\theta, \o,\phi, \phi',\psi, \psi')$ and, as in Appendix \ref{ApendiceA},
\begin{equation}
	\begin{split}
		\O(p)_n^{\m_1..\m_n}&\equiv  
		\begin{cases}
			p^{\m_1} ... p^{\m_n}~, \textrm{if $n\geq 0$} \\
			0, \textrm{if $n<0$}~~;
		\end{cases}
		\\[2ex]
		\zeta_n^{\m_1...\m_n} &\equiv
		\begin{cases}
			\eta^{\m_1 \m_2}... \eta^{\m_{n-1}\m_n} + \textrm{dist. perm.}, \textrm{ if $n$ is even,}  \\
			0, \textrm{ if $n$ is odd or $n<0$}~.
		\end{cases} 
	\end{split}
\end{equation}
Also, using (\ref{p,k}), we can rewrite $p'$ as
\begin{equation}
	{p'}^\m=Ap^\m +Bk^\m
\end{equation}
where
\begin{equation}
	A=-\frac{x_5}{1-x_2}~~,~~~~~~B=\frac{x_1+x_2+x_5+x_6-1}{1-x_2}~~.
\end{equation}
So, we have
\begin{equation}
	\O(p')_{6-2j}^{\m_1...\m_{6-2j}} = {p'}^{\m_1}\cdots {p'}^{\m_{6-2j}} = 
	\sum_{i=0}^{6-2j}A^i B^{6-2j-i}\{p_i k_{6-2j-i}\}^{\m_1...\m_{6-2j}}~~.
\end{equation}
The previous result can be used in (\ref{X'00}), and it reads
\begin{equation}
	\begin{split}
		X'_{00}=&-\eta_{\phi \phi'}\eta_{\psi \psi'} C(8) \sum_{j=0}^3 \sum_{i=0}^{6-2j} \left(-\frac{1}{2}\right)^j \G(13/2-j)(c'+{p'}^2)^{-13/2+j}\times \\
		&~~~~\times \bigg(A^i B^{6-2j-i}\{p_i k_{6-2j-i}\}^{\m_1...\m_{6-2j}}\zeta_{2j}^{\m_{6-2j+1}... \m_6}+\textrm{dist. perm.}\bigg)k^{\chi}k^2 p^\xi\eta^{\a\b} ~~. \\
		=&-\eta_{\phi \phi'}\eta_{\psi \psi'}  p^\xi C(8) \sum_{j=0}^3 \sum_{i=0}^{6-2j} \left(-\frac{1}{2}\right)^j \G(13/2-j)(c'+{p'}^2)^{-13/2+j}\times \\
		&~~~~\times \bigg(A^i B^{6-2j-i}\{p_i k_{9-2j-i}\}^{\m_1...\m_{6-2j}\bar{\chi}_k}\zeta_{2j}^{\m_{6-2j+1}... \m_6}+\textrm{dist. perm.}\bigg)\eta^{\a\b} ~~.
	\end{split}
\end{equation}

It is also necessary to integrate $X'_{00}$ with respect to the momentum $k_2$, which has been
relabeled as $k$. For this purpose, the previous result can be put into the form of a $J_r$ integral. The term $(c'+{p'}^2)^{-13/2+j}$ appears in the denominator, giving $\alpha=-13/2+j$, whereas
$k_{9-2j-i}$ appears in the numerator, and the power of $k$ is given by $9-2j-i$. In fact,
rewriting the term $(c'+{p'}^2)$ in the form $(k^2+2 k\cdot p'' -c'')$, the integration with respect to the momentum $k$ becomes straightforward and consists of replacing $k_{9-2j-i}$ by $J_{9-2j-i}((-13/2+j), p'', c'')$. Thus, $c'+{p'}^2$
can be rewritten as
\begin{equation}
	c'+{p'}^2 = \frac{a}{(1-x_2)^2}(k^2+2 k \cdot p'' -c'')~,
\end{equation}
where
\begin{equation}
\begin{split}
		a&=(x_1+x_2+x_5+x_6-1)^2+(x_1+x_5-1)(1-x_2)~, \\[2ex]
		p'' &= \frac{x_5(1-x_1-x_2-x_5-x_6)+x_6(1-x_2)}{a} p~, \\[2ex]
		c'' &= -\frac{[x_5^2-(1-x_2)(x_5+x_6)]p^2+(1-x_1-x_2-x_4-x_5-x_6)(1-x_2)\m^2}{a}.
\end{split}
\end{equation}
So, defining 
\begin{equation}
	M(j)=\left(-\frac{1}{2}\right)^j C(8)\left(\dfrac{a}{(1-x_2)^2}\right)^{-13/2+j} \G(13/2-j)~,
\end{equation}
the expression for $X'_{00}$ reads
\begin{equation}
	\begin{split}
	X'_{00} =-\eta_{\phi \phi'}\eta_{\psi \psi'}  p^\xi  \sum_{j=0}^3 \sum_{i=0}^{6-2j}  \frac{\left[M(j) A^i B^{6-2j-i}\{p_i k_{9-2j-i}\zeta_{2j}\}^{\theta \o \phi \phi' \psi \psi' \bar{\chi}_k}\right] \eta^{\a\b}}{(k^2+2 k\cdot p''-c'')^{13/2-j}}
	  ~.
	\end{split}
\end{equation}
where $\zeta$, the product of $\eta'$s, has been absorbed into the expansion operator. As can be noted, the number of indices is less than the sum of the subscripts in the expansion operator and it is due to the presence of a factor $k^2$. It should be stressed that after integrating $X'_{00}$ with respect to $k$, no divergence will be encountered. Indeed, $J_{9-2j-i}$ contains $\Gamma(\alpha'-d/2-j')$, where $\alpha'=13/2-j$ and $j'$ ranges from 0 to $\lfloor 9/2-j-i/2 \rfloor$. One may wonder whether the argument of the $\Gamma$ function could be a negative integer or not. However, it can be checked that $\Gamma(\alpha'-d/2-j')=\Gamma(5-j-j')$, and for $j=0$, the maximum value of $j'$ is 4, and as we increase $j$ by one unit, the maximum value of $j'$ decreases by one unit. Therefore, the smallest argument of the $\Gamma$ function is 1 and there are no divergences.

It remains to determine the other $X'_{0l'}$ terms. Observing the expression for $X'_0$ in (\ref{X'0}), it is possible to write the expressions for $X'_{01},~...,~X'_{0,10}$ from $X'_{00}$. For instance, $X'_{01}=-J_5^\theta g_4^{\xi \chi}p^\o$ is
\begin{equation}
	\begin{split}
		X'_{01}= &-\eta_{\phi \phi'}\eta_{\psi \psi'}  p^\o  \sum_{j=0}^2 \sum_{i=0}^{5-2j} M(j) \frac{\left[A^i B^{5-2j-i}\{p_i k_{9-2j-i}\zeta_{2j}\}^{\theta \phi \phi' \psi \psi' \bar{\chi}_k \bar{\xi}_k}\right]\eta^{\a\b}}{(k^2+2 k\cdot p''-c'')^{13/2-j}}
	\end{split}
\end{equation}
It is worth noting how the sum over $j$ changed. This was due to $X'_{01}$ having a $J_5$ instead of a $J_6$. Consequently, it also affected the sum over $i$. Moreover, the indices with an upper bar changed due to the modifications in the indices of $g$,
which is $g_4^{\xi\chi}=k^2 k^{\xi} k^{\chi}$ in this case. Using a similar procedure,
the remaining $X'_{0l'}$ are given by:
\begin{equation}
\begin{split}
	X'_{02}= &\eta_{\phi \phi'}\eta_{\psi \psi'}  p^\o p^\xi \sum_{j=0}^2 \sum_{i=0}^{5-2j} M(j) \frac{\left[A^i B^{5-2j-i}\{p_i k_{8-2j-i}\zeta_{2j}\}^{\theta \phi \phi' \psi \psi' \bar{\chi}_k }\right]\eta^{\a\b}}{(k^2+2 k\cdot p''-c'')^{13/2-j}};
\end{split}
\end{equation}
\begin{equation}
\begin{split}
		X'_{03}=&2\eta_{\phi \phi'}\eta_{\psi \psi'}  p_\d  \sum_{j=0}^3 \sum_{i=0}^{6-2j} M(j) \frac{\left[A^i B^{6-2j-i}\{p_i k_{9-2j-i}\zeta_{2j}\}^{\theta \o \phi \phi' \psi \psi' \bar{\chi}_k\bar{\xi}_k \bar{\d}_k}\right]\eta^{\a\b}}{(k^2+2 k\cdot p''-c'')^{13/2-j}};
\end{split}
\end{equation}
\begin{equation}
\begin{split}
	X'_{04} =& -\eta_{\phi \phi'}\eta_{\psi \psi'}  p^2  \sum_{j=0}^3 \sum_{i=0}^{6-2j} M(j)\frac{\left[A^i B^{6-2j-i}\{p_i k_{8-2j-i}\zeta_{2j}\}^{\theta \o \phi \phi' \psi \psi' \bar{\xi}_k \bar{\chi}_k}\right]\eta^{\a\b}}{(k^2+2 k\cdot p''-c'')^{13/2-j}};
\end{split}
\end{equation}
\begin{equation}
\begin{split}
	X'_{05} =& 2\eta_{\phi \phi'} p_\d  \sum_{j=0}^2 \sum_{i=0}^{5-2j} M(j) \frac{\left[A^i B^{5-2j-i}\{p_i k_{9-2j-i}\zeta_{2j}\}^{\theta \phi \phi'  \o \d \bar{\chi}_k \bar{\xi}_k}\right]\eta^{\a\b}}{(k^2+2 k\cdot p''-c'')^{13/2-j}}; \\[2ex]
\end{split}
\end{equation}
\begin{equation}
	\begin{split}
		X'_{06} =& -4 \eta_{\phi \phi'}  p_\d p_{\d'} \sum_{j=0}^2 \sum_{i=0}^{5-2j} M(j) \frac{\left[A^i B^{5-2j-i}\{p_i k_{8-2j-i}\zeta_{2j}\}^{\theta \o \phi \phi' \d  \bar{\chi}_k \bar{\xi}_k \bar{\d'}_k }\right]\eta^{\a\b}}{(k^2+2 k\cdot p''-c'')^{13/2-j}};
	\end{split}
\end{equation}
\begin{equation}
	\begin{split}
		X'_{07} =&2\eta_{\phi \phi'} p^2 p_\d  \sum_{j=0}^2 \sum_{i=0}^{5-2j} M(j) \frac{\left[A^i B^{5-2j-i}\{p_i k_{7-2j-i}\zeta_{2j}\}^{\theta \phi \phi'  \o \d \bar{\chi}_k \bar{\xi}_k}\right]\eta^{\a\b}}{(k^2+2 k\cdot p''-c'')^{13/2-j}};\\[3ex]
	\end{split}
\end{equation}
\begin{equation}
	\begin{split}
		X'_{08} =&-\eta_{\phi \phi'} p^2  \sum_{j=0}^2 \sum_{i=0}^{4-2j} M(j) \frac{\left[A^i B^{4-2j-i}\{p_i k_{8-2j-i}\zeta_{2j}\}^{\theta \o \phi \phi' \bar{\chi}_k \bar{\xi}_k}\right]\eta^{\a\b}}{(k^2+2 k\cdot p''-c'')^{13/2-j}};
		\\[3ex]
	\end{split}
\end{equation}
\begin{equation}
	\begin{split}
		X'_{09} =&2\eta_{\phi \phi'} p^2 p_\d \sum_{j=0}^2 \sum_{i=0}^{4-2j} M(j) \frac{\left[A^i B^{4-2j-i}\{p_i k_{7-2j-i}\zeta_{2j}\}^{\theta \o \phi \phi' \bar{\chi}_k \bar{\xi}_k}\bar{\d}_k\right]\eta^{\a\b}}{(k^2+2 k\cdot p''-c'')^{13/2-j}};
	\end{split}
\end{equation}
\begin{equation}
	\begin{split}
		X'_{0,10} =&-\eta_{\phi\phi'}p^2 p^2  \sum_{j=0}^1 \sum_{i=0}^{2-2j} M(j) \frac{\left[A^i B^{2-2j-i}\{p_i k_{6-2j-i}\zeta_{2j}\}^{\theta \o \phi\phi'  \bar{\chi}_k \bar{\xi}_k}\right]\eta^{\a\b}}{(k^2+2 k\cdot p''-c'')^{13/2-j}};
	\end{split}
\end{equation}
All of the $X'_{0l'}$ will lead to non-divergent and dimensionless terms when integrated with respect to $k$.

The previous calculations account only for the contributions of $X'_0$. It is still necessary to determine the contributions for the remaining
$X'_l=\displaystyle \int \frac{d^3 k}{(2\pi)^3} \frac{X_l}{(k_1^2+2 k_1 \cdot p' -c')^8}$. This can be accomplished by using equation (\ref{X}), and we have detailed this procedure in Appendix \ref{ApendiceC}. Thus, after integrating each $X'_l$ with respect
to $k$, we obtain
\begin{equation}
	X''_{l} = \int \frac{d^3 k}{(2\pi)^3} X'_{l} = \sum_{l'=0}^{10}\int \frac{d^3 k}{(2\pi)^3} X'_{ll'} = \sum_{l'=0}^{10} X''_{ll'},
\end{equation}
where $X''_{ll'}$ is obtained from $X'_{ll'}$ just by eliminating the denominator and replacing $k_r$ by $J_r(13/2-j,p'',c'')$, because
\begin{equation}
	\int \frac{d^3 k}{(2\pi)^3}\frac{\{p_{n_1} k_{n_2} \zeta_{n_3} \}^{\m_1 ...}}{(k^2+2k\cdot p'' - c'')^{13/2-j}} =
	\bigg\{p_{n_1} J_{n_2}\left(13/2-j,p'',c''\right) \zeta_{n_3} \bigg\}^{\m_1 ...}~.
\end{equation}
Therefore, the renormalized diagram $\g_{8\pm}$ yields
\begin{equation}
	\int \frac{d^3 k_1}{(2\pi)^3} \frac{d^3 k_2}{(2\pi)^3} R^{\m\n}_{\g_{8_\pm}}
	=\sum_{l=0}^6 \sum_{l'=0}^{10}\int d\O X''_{ll'}~.
\end{equation}

The integrals in terms of $d\O$, which arise from the Feynman parametrization, yield just a numerical factor. Furthermore, all of them produce non-local terms, since all $X''_{ll'}$ contain external momenta in the denominators. To prove our last statement, it is necessary to note that all the terms of each
$X''_{ll'}$ contain a $J_{n-2j-i}(13/2,p'',c'')$, which can be expressed as \footnote{$i$ and $j$ in $J_{n-2j-i}$ are the sum indices that appeared in the expressions of the terms $X'_{ll'}$.}
\begin{equation}
  J_{n-2j-i}(13/2,p'',c'')= \sum_{j'=0}^{\left\lfloor \frac{n-2j-i}{2} \right\rfloor} f(j') (c''+{p''}^2)^{3/2-13/2+j+j'}~,
\end{equation}
where $f(j')$ includes all the terms that are not important in our analysis here and $n$ varies from 6 to 9, as can be seen in the expressions of $X''_{ll'}$.
The power of \linebreak
$(c''+{p''}^2)$ is $3/2-13/2+j+j'=-5+j+j'$ and we can prove that it is
always negative. The highest value of $j'$ is attained when $i=0$ and $n=9$.
In this scenario, \linebreak
$j'=\left\lfloor \frac{9-2j}{2} \right\rfloor= \left\lfloor \frac{8-2j}{2}\right\rfloor= 4-j$. Consequently, the lowest power of $(c''+{p''}^2)$
is 
\linebreak
$-5+j+(4-j)=-1$, regardless any value of $j$ and we can conclude that all the $X''_{ll'}$ yield only non-local terms. 
Therefore, the 2-loops vacuum polarizations do not contribute to the renormalization of the gauge field mass $\mu$.

In summary, by taking into account all the 1- and 2-loops graphs, convergent or finite by UV and IR subtractions, it can be concluded that the $\beta$-functions associated to the coupling constants $e$ (electric charge) and $g$ (pseudochiral charge), and the gauge field mass $\mu$, are vanishing up 2-loops and beyond, since the parity-preserving massless $U(1)\times U(1)$ Maxwell-Chern-Simons QED$_3$ model \cite{DeLima} is superrenormalizable and has divergent diagrams only up to order of 2-loops. Also, as discussed in \cite{parity}, together with the explicitly computations presented here, the parity symmetry is preserved by BPHZL renormalization method for the model \cite{DeLima}.

\section{Conclusion}

The parity-preserving massless $U(1)\times U(1)$ Maxwell-Chern-Simons QED$_3$ model \cite{DeLima} exhibits 
ultraviolet and infrared quantum finiteness, all the physical parameters --  
the coupling constants $e$ (electric charge) and $g$ (pseudochiral charge), and the gauge field mass $\mu$ -- do not get renormalized, thus their respective $\beta$-functions vanish. Moreover, the model shows parity conservation at any loop order despite the infrared divergence subtractions. The proof was done by using the Bogoliubov-Parasiuk-Hepp-Zimmermann-Lowenstein (BPHZL) renormalization method, nevertheless thanks to the presence of massless fermions, infrared divergences shall be induced by the ultraviolet divergence subtractions, for this reason, the Lowenstein-Zimmermann (LZ) subtraction scheme was adopted. The ultraviolet and infrared power-counting formula \ref{power_counting} implies that ultraviolet divergences are bounded by two loops, beyond that all 1-particle irreducible Feynman diagrams are finite due to the superrenormalizability of the model.

At 1-loop order, all six vacuum-polarization tensor graphs are ultraviolet linear divergent, four of the six self-energy are ultraviolet logarithm divergent, and all the vertex-function diagrams are ultraviolet finite (Fig. 1 and Table 2). Moreover at 2-loops order, twenty four of the thirty six vacuum-polarization tensor Feynman graphs are ultraviolet divergent (Fig. 2). 

The BPHZL subtraction scheme was defined, the 1-loop polarization tensors and self-energies presented, and the expressions for the ultraviolet divergent graphs at 2-loops were written. The Zimmermann's forests were built up for the BPHZL renormalization at 2-loops of the logarithmically divergent vacuum polarization graphs $\g_{11_\pm}$, $\g_{12_\pm}$, $\g_{8_\pm}$ and 
$\g_{9_\pm}$, for further renormalization of those graphs based on the relations displayed in the appendices, where it should be stressed that, as far as we know, the general expression for the integral $J_r^{\m_1 ... \m_r}(\a,p,c)$ (B.7) is new in the literature. 

As last comments, opposite to the case of the ordinary massless $U(1)$ QED$_3$, where the BPHZL method breaks parity \cite{bphzl}, for the model studied here \cite{DeLima}, parity is preserved by the BPHZL subtraction method with the Lowenstein's adaptation of the Zimmermann's forest formula. Finally, since the vertex graphs $\g_{4_\pm}$ are convergent, and all renormalized vacuum polarization graphs $\g_{11_\pm}$, $\g_{12_\pm}$, $\g_{8_\pm}$ and $\g_{9_\pm}$ yields only nonlocal counterterms, consequently the $\beta$-functions associated to the gauge coupling constants and mass vanish.

	\acknowledgments
	
	The authors thank CAPES-Brazil for invaluable financial help.

\appendix
\section{Some notations and useful relations}\label{ApendiceA}
\quad It is important to establish a notation that facilitates calculations. We define two tensors, $\O$ and $\zeta$, where the former represents the product of
momenta, whereas the latter represents the product of the metric tensor $\eta$ :
\begin{equation}
	\begin{split}
		\O(p)_n^{\m_1..\m_n}&\equiv  
		\begin{cases}
			p^{\m_1} ... p^{\m_n}~, \textrm{if $n\geq 0$} \\
			0, \textrm{if $n<0$}~~;
		\end{cases}
		\\[2ex]
		\zeta_n^{\m_1...\m_n} &\equiv
		\begin{cases}
			\eta^{\m_1 \m_2}... \eta^{\m_{n-1}\m_n} + \textrm{dist. perm.}, \textrm{ if $n$ is even,}  \\
			0, \textrm{ if $n$ is odd or $n<0$}~.
		\end{cases} 
	\end{split}
\end{equation}
Both tensors are completely symmetric, and by "dist. perm." we mean all distinct permutations. For instance, for the tensor $\zeta$, permutations between indices on the same $\eta$ are disregarded. In addition, if "dist. perm." is used after a sum
of two or more tensors, then permutations between indices of different terms are forbidden. Nevertheless, permutations between indices of a product are allowed.
As a final comment about the distinct permutations, if a tensor index has a bar, then
it must be fixed and does not participate in the permutations. For example,
in the expression $A^{\bar{\r}\m\n\s}B^{\o\d}+\textrm{dist. perm.}$, the index $\r$ cannot be permuted, and terms such as $B^{\r\a}$, where $\a$ is an arbitrary index, are not allowed. 

It can be shown that $\zeta$ obeys the following relation:
\begin{equation}\label{zeta2}
	\zeta_{2n+2}^{\mu_1...\mu_{2n+2}}= \zeta_{2n}^{\mu_1...\mu_{2n}}\eta^{\mu_{2n+1}\mu_{2n+2}} + \textrm{dist. perm.}
\end{equation}
Indeed, suppose $\eta^{\nu_1 \nu_2}... \eta^{\nu_{2n+1} \nu_{2n+2}}$ is a term in $\zeta_{2n}^{\mu_1...\mu_{2n+2}}$, where $(\nu_1,...,\nu_{2n+2})$ is a permutation of $(\mu_1,...,\mu_{2n+2})$. Since $\eta^{\nu_1 \nu_2}... \eta^{\nu_{2n+1} \nu_{2n+2}}$ is also a term of $\zeta_{2n}^{\n_1...\n_{2n}}\eta^{\n_{2n+1}\n_{2n+2}}$, and the latter is a term of the right-hand side of (\ref{zeta2}), then every term of $\zeta_{2n+2}^{\mu_1...\mu_{2n+2}}$ is a term of $(\zeta_{2n}^{\mu_1...\mu_{2n}}\eta^{\mu_{2n+1}\mu_{2n+2}} + \textrm{dist. perm.})$. Conversely, by the definition of $\zeta$, every term of $(\zeta_{2n}^{\mu_1...\mu_{2n}}\eta^{\mu_{2n+1}\mu_{2n+2}} + \textrm{dist. perm.})$ is a term of $\zeta_{2n+2}^{\mu_1...\mu_{2n+2}}$. Therefore, we conclude that (\ref{zeta2}) is true.

Another property of the tensors $\O$ and $\zeta$ is
\begin{equation}\label{important_derivative}
	\frac{\partial}{\partial p_\rho}\bigg(\O_n^{\mu_1...\mu_n}\zeta_{2m}^{\mu_{n+1}...\mu_{n+2m}}+\textrm{dist. perm}\bigg)
	=\O_{n-1}^{\mu_1...\mu_{n-1}}\zeta_{2m+2}^{\bar{\rho}\mu_{n}...\mu_{n+2m}}+\textrm{dist. perm}
\end{equation}
We can prove the previous result by induction. If $m=0$, then
\begin{equation}
	\frac{\partial}{\partial p_\rho}\O_n^{\mu_1...\mu_n} = \O_{n-1}^{\mu_1...\mu_{n-1}}\zeta_2^{\bar{\rho}\mu_n} + \textrm{dist. perm.},
\end{equation}
where the distinct permutations are between the second index of $\zeta_2$ and the
indices of $\O_{n-1}$.
Let us assume that (\ref{important_derivative}) is true. Multiplying both sides of (\ref{important_derivative}) by $\eta^{\mu_{n+2m+1}\mu_{n+2(m+1)}}$, considering all the distinct permutations between $\zeta$ and $\eta$, and using (\ref{zeta2}), we have
\begin{equation}
	\frac{\partial}{\partial p_\rho}\bigg(\O_n^{\mu_1...\mu_n}\zeta_{2(m+1)}^{\mu_{n+1}...\mu_{n+2(m+1)}}+\textrm{dist. perm}\bigg)
	=\O_{n-1}^{\mu_1...\mu_{n-1}}\zeta_{2(m+1)+2}^{\bar{\rho}\mu_{n}...\mu_{n+2(m+1)}}+\textrm{dist. perm}~,
\end{equation}
which proves that (\ref{important_derivative}) is valid.

\section{The J$_r$-integrals}\label{ApendiceB}

\quad Here are presented the J$_r$-integrals, frequently encountered in loop calculations. Also, some useful and significant results are shown. These results remain applicable across arbitrary dimensions, whether integers or not. A general J$_r$-integral expression is
\begin{equation}
	J_r^{\m_1...\m_r}(\a,p,c) = \int \frac{d^D k}{(2\pi)^D} \frac{k^{\m_1}... k^{\m_r}}{(k^2+2p\cdot k - c)^\a}\,.
\end{equation}
The three subsequent results can be found in \cite[pp. 83]{Collins}.
\begin{equation}\label{J0}
	J_0\equiv \int \frac{d^d k}{(2 \pi)^d}\frac{1}{(k^2+2pk-c)^\a}
	= i(-1)^\a\frac{\pi^{d/2}}{(2\pi)^d}(c+p^2)^{d/2-\a}\frac{\G(\a-d/2)}{\G(\a)}~.~~~~~
\end{equation}
\begin{align}\label{J1}
	J_1^\m\equiv \int \frac{d^d k}{(2 \pi)^d}\frac{1}{(k^2+2pk-c)^\a}
	= i(-1)^{\a+1}\frac{\pi^{d/2}}{(2\pi)^d}(c+p^2)^{d/2-\a}\,p^\m\frac{\G(\a-d/2)}{\G(\a)}~.
\end{align}
\begin{align}\label{J2}
	\nonumber
	J_2^{\m\n}\equiv \int \frac{d^d k}{(2 \pi)^d}&\frac{k^\m k^\n}{(k^2+2pk-c)^\a}
	= i(-1)^{\a}\frac{\pi^{d/2}}{(2\pi)^d}(c+p^2)^{d/2-\a}\times~~~~~~~~ ~~\\
	&\times \left[\frac{\G(\a-d/2)p^\m p^\n-\frac{1}{2}\G(\a-1-d/2)\eta^{\m\n}(c+p^2)}{\G(\a)}\right]~.~~~~~~~~~~~~
\end{align}

It is important to note that there is a recurrence relation from where $J_{n+1}^{\m_1\cdots \m_{n+1}}$ can be obtained,
using the derivative of $J_{n}^{\m_1\cdots \m_{n}}$ in respect to $p^{\m_{n+1}}$. The recurrence relation is
\begin{equation}\label{recurrencerelation}
	J_{r+1}^{\m_1...\m_{r+1}}(\a,p,c)=-\frac{1}{2}~\frac{1}{\a-1}~\frac{\pa J^{\m_1... \m_r}_r }{\pa p_{\m_{r+1}}}(\a-1,p,c)~.
\end{equation}
Thus, the expression for $J_3$ is
\begin{align}\label{J3}
	\nonumber
	& J_3^{\m\n\b}\equiv \int  \frac{d^d k}{(2 \pi)^d}\frac{k^\m k^\n k^\b}{(k^2+2pk-c)^\a}
	= i(-1)^{\a}\frac{\pi^{d/2}}{(2\pi)^d}(c+p^2)^{d/2-\a}\times~~~~~~~~ ~~\\
	&\times \left[\frac{\G(\a-d/2)p^\m p^\n p^\b-\frac{1}{2}\G(\a-1-d/2)\left(\eta^{\m\n}p^\b+\eta^{\n\b}p^\m+\eta^{\b\m}p^\n\right)(c+p^2)}{\G(\a)}\right]~.
\end{align}

There is a general expression\footnote{To the best of our knowledge, it is not found in the literature.} for the results of these integrals:
\begin{equation}\label{generalJ}
	\begin{split}
		J_r^{\m_1 ... \m_r}(\a,p,c)&=i(-1)^{r+\a} \frac{\pi^{D/2}}{\G(\a)(2\pi)^D}\sum_{j=0}^{\lfloor \frac{r}{2} \rfloor}\left(-\frac{1}{2}\right)^j \G(\a-d/2-j) \times\\
		&\times (c+p^2)^{d/2-\a+j}\bigg(\O(p)_{r-2j}^{\m_1 ... \m_{r-2j}} \zeta_{2j}^{\m_{r-2j+1}... \m_r}+\textrm{dist. perm.}\bigg)~~.
	\end{split}
\end{equation}
where $\lfloor\frac{r}{2}\rfloor$ is the floor function, which returns the closest smaller integer of $\frac{r}{2}$. The previous expression is particularly useful when dealing with J$_r$-integrals with $r$ bigger than 3. It can be shown the validity of (\ref{generalJ}) by using the inductive method. As can be checked, it is valid for $J_0$. Let us suppose it is valid for an arbitrary $r$. Differentiating $J_r$ with respect to $p_{\m_{r+1}}$ we have
\begin{equation}\label{A8}
	\begin{split}
		\frac{\pa J_r^{\m_1...\m_r}}{\pa p_{\m_{r+1}}}&(\a,p,c)=i(-1)^{r+\a} \frac{\pi^{D/2}}{\G(\a)(2\pi)^D}\sum_{j=0}^{\lfloor \frac{r}{2} \rfloor}\left(-\frac{1}{2}\right)^j \G(\a-d/2-j) \times\\
		&\times\frac{\pa}{\pa p_{\m_{r+1}}}\bigg[ (c+p^2)^{d/2-\a+j}\bigg(\O(p)_{r-2j}^{\m_1 ... \m_{r-2j}} \zeta_{2j}^{\m_{r-2j+1}... \m_r}+\textrm{dist. perm.}\bigg)\bigg]~~. 
	\end{split}
\end{equation}
Taking into account (\ref{important_derivative}) and
\begin{equation}
	p^{\m_{r+1}}\bigg(\O(p)_{r-2j}^{\m_1...\m_{r-2j}}\zeta_{2j}^{\m_{r-2j+1}... \m_r}+\textrm{dist. perm.}\bigg)=\O(p)_{r-2j+1}^{\bar{\m}_{r+1}\m_1 ... \m_{r-2j}}
	\zeta_{2j}^{\m_{r-2j+1}... \m_r}+\textrm{dist. perm.}~,
\end{equation}
the derivative in (\ref{A8}) can be expanded as
\begin{equation}
	\begin{split}
		&\frac{\pa}{\pa p_{\m_{r+1}}}\bigg[ (c+p^2)^{d/2-\a+j}\bigg(\O(p)_{r-2j}^{\m_1 ... \m_{r-2j}} \zeta_{2j}^{\m_{r-2j+1}... \m_r}+\textrm{dist. perm.}\bigg)\bigg] = \\[2ex]
		&	= 2(d/2-\a+j)(c+p^2)^{d/2-(\a+1)+j}\bigg(\O(p)_{r-2j+1}^{\bar{\m}_{r+1}\m_1 ... \m_{r-2j}} \zeta_{2j}^{\m_{r-2j+1}... \m_r}+\textrm{dist. perm.}\bigg)+ \\[2ex]
		& +(c+p^2)^{d/2-\a+j} 
		\bigg(\O(p)_{r-2j-1}^{\m_1 ... \m_{r-2j-1}} \zeta_{2(j+1)}^{\bar{\m}_{r+1}{\m}_{r-2j}... \m_{r}}+\textrm{dist. perm.}\bigg)~,
	\end{split}~
\end{equation}
where indices with an upper bar do not permute. Furthermore, since\linebreak
$\G(n+1)=n\G(n)$, then
\begin{equation}\label{A10}
	\begin{split}
		\sum_{j=0}^{\lfloor \frac{r}{2} \rfloor}&\left(-\frac{1}{2}\right)^j \G(\a-d/2-j)
		~~2(d/2-\a+j)(c+p^2)^{d/2-(\a+1)+j}\times \\[2ex]
		\times &\bigg(\O(p)_{r-2j+1}^{\bar{\m}_{r+1}\m_1 ... \m_{r-2j}} \zeta_{2j}^{\m_{r-2j+1}... \m_r}+\textrm{dist. perm.}\bigg) = \\
		\sum_{j=0}^{\lfloor \frac{r}{2} \rfloor}&\left(-\frac{1}{2}\right)^{j-1} \G(\a+1-d/2-j)
		~~(c+p^2)^{d/2-(\a+1)+j}\times \\[2ex]
		\times &\bigg(\O(p)_{r-2j+1}^{\bar{\m}_{r+1}\m_1 ... \m_{r-2j}} \zeta_{2j}^{\m_{r-2j+1}... \m_r}+\textrm{dist. perm.}\bigg)~~,
	\end{split}
\end{equation}
where $2(d/2-\a+j)$ have been written as $-2(\a-d/2-j)$ and absorbed in $-(1/2)^j \G(\a-d/2-j)$. In addition, by performing the change $j\rightarrow j-1$, the next
summation can be rewritten in the following way:
\begin{equation}
	\begin{split}
		\sum_{j=0}^{\lfloor \frac{r}{2} \rfloor}&\left(-\frac{1}{2}\right)^j \G(\a-d/2-j)
		(c+p^2)^{d/2-\a+j} 
		\bigg(\O(p)_{r-2j-1}^{\m_1 ... \m_{r-2j-1}} \zeta_{2(j+1)}^{\bar{\m}_{r+1}{\m}_{r-2j}... \m_{r}}+\textrm{dist. perm.}\bigg)= \\[2ex]
		&=\sum_{j=1}^{\lfloor \frac{r}{2} \rfloor +1}\left(-\frac{1}{2}\right)^{j-1} \G(\a+1-d/2-j)
		(c+p^2)^{d/2-(\a+1)+j} 
		\\[2ex]
		&~~~~~~~~~~~~~~~~~~~~~~\bigg(\O(p)_{r-2j+1}^{\m_1 ... \m_{r-2j+1}} \zeta_{2j}^{\bar{\m}_{r+1}{\m}_{r-2j+2}... \m_{r}}+\textrm{dist. perm.}\bigg)~~.
	\end{split}
\end{equation}
It is worth noting that in the aforementioned expression, if $r$ is even, the term with
$j=\lfloor r/2 \rfloor+1$ vanishes, since it comes from the differentiation, with respect
to $p_{\m_{r+1}}$, of a term that does not depend on $p$ ($\O_0 \zeta_{2r} $). So,
if $r$ is even, the summation
must encompass terms up to $\lfloor r/2 \rfloor =\lfloor (r+1)/2 \rfloor$. The last equality is due to $r$ being even. Moreover, if $r$ is odd, then $\lfloor r/2 \rfloor+1=\lfloor (r+1)/2 \rfloor$ and in both cases we can change $\sum_{j=1}^{\lfloor \frac{r}{2} \rfloor +1}$ for $\sum_{j=1}^{\lfloor \frac{r+1}{2} \rfloor}$. We can also perform, in the summation present in (\ref{A10}), the change
$\sum_{j=0}^{\lfloor \frac{r}{2} \rfloor}$ for $\sum_{j=0}^{\lfloor \frac{r}{2} \rfloor +1}$
since $\O_{-1}=0$, and then we perform the change $\sum_{j=1}^{\lfloor \frac{r}{2} \rfloor +1}$ for $\sum_{j=1}^{\lfloor \frac{r+1}{2} \rfloor}$. Therefore, making the term for $j=0$ in (\ref{A10}) explicit, (\ref{A8}) reads
\begin{equation}\label{B13}
	\begin{split}
		&\frac{\pa J_r^{\m_1...\m_r}}{\pa p_{\m_{r+1}}}(\a,p,c)=i(-1)^{r+\a} \frac{\pi^{D/2}}{\G(\a)(2\pi)^D}\bigg[\bigg(-2 \G(\a+1-d/2)(c+p^2)^{d/2-(\a+1)}\times\\[2ex]
		&\times \O(p)_{r+1}^{\m_1...\m_{r+1}}\bigg)+
		\sum_{j=1}^{\lfloor \frac{r+1}{2} \rfloor} \left(-\frac{1}{2}\right)^{j-1} \G(\a+1-d/2-j)
		(c+p^2)^{d/2-(\a+1)+j} \times \\[2ex]
		&\times\bigg(\O(p)_{r-2j+1}^{\bar{\m}_{r+1}\m_1 ... \m_{r-2j}} \zeta_{2j}^{\m_{r-2j+1}... \m_r}+\O(p)_{r-2j+1}^{\m_1 ... \m_{r-2j+1}} \zeta_{2j}^{\bar{\m}_{r+1}{\m}_{r-2j+2}... \m_{r}}+\textrm{dist. perm.}\bigg)\bigg]~.
	\end{split}
\end{equation}
In the last term of the previous equation, the missing permutations between
the index with an upper bar were obtained, since the referred index appear in both tensors, $\O$ and $\zeta$.
Then, (\ref{B13}) reads
\begin{equation}
	\begin{split}
		&\frac{\pa J_r^{\m_1...\m_r}}{\pa p_{\m_{r+1}}}(\a,p,c)=i(-1)^{r+\a} \frac{\pi^{D/2}}{\G(\a)(2\pi)^D}\bigg[\bigg(-2 \G(\a+1-d/2)(c+p^2)^{d/2-(\a+1)}\times\\[2ex]
		&\times \O(p)_{r+1}^{\m_1...\m_{r+1}}\bigg)+
		\sum_{j=1}^{\lfloor \frac{r+1}{2} \rfloor} \left(-\frac{1}{2}\right)^{j-1} \G(\a+1-d/2-j)
		(c+p^2)^{d/2-(\a+1)+j} \times \\[2ex]
		&\times\bigg(\O(p)_{r-2j+1}^{\m_1 ... \m_{r-2j+1}} \zeta_{2j}^{{\m}_{r+1}{\m}_{r-2j+2}... \m_{r}}+\textrm{dist. perm.}\bigg)\bigg]~.
	\end{split}
\end{equation}
Thus, by the recurrence relation (\ref{recurrencerelation}), in the last expression, multiplying all the terms by $-1/(2(\a+1))$ and then performing the
change $\a\rightarrow \a-1$, and
due to the fact that $(-1)^{r+a-1}=(-1)^{r+a+1}$, we obtain
\begin{equation}
	\begin{split}
		J^{\m_1...\m_{r+1}}_{r+1}(\a,p,c)	&=i(-1)^{(r+1)+\a} \frac{\pi^{D/2}}{\G(\a)(2\pi)^D}\sum_{j=0}^{\lfloor \frac{r+1}{2} \rfloor}\left(-\frac{1}{2}\right)^j \G(\a-d/2-j) \times\\
		&\times (c+p^2)^{d/2-\a+j}\bigg(\O(p)_{(r+1)-2j}^{\m_1 ... \m_{(r+1)-2j}} \zeta_{2j}^{\m_{(r+1)-2j+1}... \m_{(r+1)}}+\textrm{dist. perm.}\bigg)~~.
	\end{split}
\end{equation}
Therefore, it can be concluded that (\ref{generalJ}) is valid.

Another important and useful result that has to be mentioned is regarding the Feynman parametrization. The denominator of a general loop integral is usually not in the form $(k^2+2p\cdot k -c)$, but it can be brought into this form by using the Feynman parametrization, a method that consists of introducing some Feynman parameters and substituting the original expression with an integral in these parameters. The simplest case of the Feynman parametrization is the following:
\begin{equation}\label{pfeynman}
	\frac{1}{\a(\a-\b)}=\int_0^1 d^3x \frac{1}{(\a-\b x)^2}~,
\end{equation}
The general case is \cite{Collins}:
\begin{equation}\label{pfeynmangeral}
	\frac{1}{A_1^{\a_1} A_2^{\a_2}\cdots A_n^{\a_n}}=
	\frac{\G\left(\sum_{i} \a_i\right)}{\prod_{j}\G(\a_j)}
	\int_0^1 dx_1 \cdots dx_n \frac{\d\left(1-\sum_{i} x_i\right)\prod_{j} x_j^{\a_j-1}}{\left(\sum_{k}A_k x_k\right)^{\sum_{l} \a_l}}~.
\end{equation}

Beyond all the results established previously, the following results, found in \cite{Gradshteyn}, are also necessary:
\begin{equation}\label{integ1}
	\int_0^1 dx ~\sqrt{x^2-x}= i\frac{\pi}{8}~,~~~~~\int_0^1 dx ~\frac{1}{\sqrt{x^2-x}}= -i\pi~.~~~~~\int_0^1 dx ~\frac{x^2}{\sqrt{x^2-x}}=-i\frac{3\pi}{8}
\end{equation}
Moreover, since $\displaystyle \frac{d}{dx}\sqrt{x^2-x}=\frac{x}{\sqrt{x^2-x}}-\frac{1}{2}\frac{1}{\sqrt{x^2-x}}$, 
it follows that
\begin{equation}\label{integ2}
	\int_0^1 dx \frac{x}{\sqrt{x^2-x}}=-\frac{i\pi}{2}~.
\end{equation}

\section{Determination of $X'_l$}\label{ApendiceC}
\quad\,
We start from $X_0$ to determine $X_1$ and $X_2$.
From (\ref{X}), we have
\begin{equation}
	X_1= -\frac{f_1^{\d'} g_{1\d'}}{f_2} X_0~~, ~~~~~~~X_2= \frac{g_2}{f_2} X_0~~.
\end{equation}
Also, considering the definition
\begin{equation}
	X'_0= \int \frac{d^3 k_1}{(2\pi)^3} \frac{X_0}{(k_1^2 + 2 k_1 \cdot p' -c')^8}
	=\sum_{l'=0}^{10} X'_{0l'}~,
\end{equation}
it was possible to obtain the results for $X'_{0l'}$. In a similar way, we can define
\begin{equation}
	X'_l= \int \frac{d^3 k_1}{(2\pi)^3} \frac{X_l}{(k_1^2 + 2 k_1 \cdot p' -c')^8}
	=\sum_{l'=0}^{10} X'_{ll'}~.
\end{equation}

Let us determine $X_1$. It has one less $k_1$ and one more $k_2$ than $X_0$. When
integrated with respect to $k_1$, each of the terms $X'_{1l'}$ will yield a $J_r$ with a subscript one unit smaller. However, the total power of the internal momenta is not changed. Therefore, to obtain $X'_{1l'}$ from $X'_{0l'}$, some minor changes in the summations and indices are needed. For example, let us check the structure of
$X'_{10}$:
\begin{equation}
	\begin{split}
		X'_{10} = &\eta_{\phi \phi'}  p^\xi  \sum_{j=0}^2 \sum_{i=0}^{5-2j} M(j) \frac{\left[A^i B^{5-2j-i}\{p_i k_{9-2j-i}\zeta_{2j}\}^{\theta \o \phi \phi' \d' \bar{\chi}_k}_{~~~~~~~~~~\bar{\d'}_k}\right]\eta^{\a\b}}{(k^2+2 k\cdot p''-c'')^{13/2-j}}~.
	\end{split}
\end{equation}
In this example, the contraction between the indices $\delta'$ in $f_1^{\delta'} g_{1\delta'}$, after integration with respect to $k_1$, gives rise to a contravariant index $\delta'$, which can be permuted, and a covariant index $\delta'$ that is not permuted and is due to a factor $k_{\delta'}$ present in all terms. Moreover, unlike what happened with $X'_{00}$, the summation over $j$ is up to $2$ instead of $3$, because it comes from a $J_5$ and not from a $J_6$. This is also responsible for the change in the summation over $i$ and the power of $B$.
In a similar way, it is possible to determine the remaining $X'_{1l'}$:
\begin{equation}
	\begin{split}
		X'_{11}= &\eta_{\phi \phi'}  p^\o  \sum_{j=0}^2 \sum_{i=0}^{4-2j} M(j)\frac{\left[A^i B^{4-2j-i}\{p_i k_{9-2j-i}\zeta_{2j}\}^{\theta \phi \phi' \d' \bar{\chi}_k \bar{\xi}_k}_{~~~~~~~~~~\bar{\d'}_k}\right]\eta^{\a\b}}{(k^2+2 k\cdot p''-c'')^{13/2-j}}~~;
	\end{split}
\end{equation}
\begin{equation}
	\begin{split}
		X'_{12}= &-\eta_{\phi \phi'}  p^\o p^\xi \sum_{j=0}^2 \sum_{i=0}^{4-2j} M(j) \frac{\left[A^i B^{4-2j-i}\{p_i k_{8-2j-i}\zeta_{2j}\}^{\theta \phi \phi' \d' \bar{\chi}_k }_{~~~~~~~~\bar{\d'}_k}\right]\eta^{\a\b}}{(k^2+2 k\cdot p''-c'')^{13/2-j}}~;
	\end{split}
\end{equation}
\\
\begin{equation}
	\begin{split}
		X'_{13}= &-2\eta_{\phi \phi'}  p_{\d} \sum_{j=0}^2 \sum_{i=0}^{5-2j} M(j) \frac{\left[A^i B^{5-2j-i}\{p_i k_{8-2j-i}\zeta_{2j}\}^{\theta \o \phi \phi' \d' \bar{\chi}_k \bar{\xi}_k \bar{\d}_k }_{~~~~~~~~~~~~~~\bar{\d'}_k}\right]\eta^{\a\b}}{(k^2+2 k\cdot p''-c'')^{13/2-j}}~~; 
	\end{split}
\end{equation}
\\
\begin{equation}
	\begin{split}
		X'_{14}= &-\eta_{\phi \phi'}  p^2 \sum_{j=0}^2 \sum_{i=0}^{5-2j} M(j) \frac{\left[A^i B^{5-2j-i}\{p_i k_{8-2j-i}\zeta_{2j}\}^{\theta \o \phi \phi' \d' \bar{\chi}_k \bar{\xi}_k  }_{~~~~~~~~~~~~\bar{\d'}_k}\right]\eta^{\a\b}}{(k^2+2 k\cdot p''-c'')^{13/2-j}}~~; 
	\end{split}
\end{equation}
\\
\begin{equation}
	\begin{split}
		X'_{15}= & -2 p_\d  \sum_{j=0}^2 \sum_{i=0}^{4-2j} M(j) \frac{\left[A^i B^{4-2j-i}\{p_i k_{9-2j-i}\zeta_{2j}\}^{\theta \o \d \d' \bar{\chi}_k \bar{\xi}_k}_{~~~~~~~~~\bar{\d'}_k}\right]\eta^{\a\b}}{(k^2+2 k\cdot p''-c'')^{13/2-j}}~; 
	\end{split}
\end{equation}
\\
\begin{equation}
	\begin{split}
		X'_{16} =& 4   p_\d p_{\d''} \sum_{j=0}^2 \sum_{i=0}^{4-2j} M(j) \frac{\left[A^i B^{4-2j-i}\{p_i k_{8-2j-i}\zeta_{2j}\}^{\theta \o  \d \d' \bar{\d''_k} \bar{\chi}_k \bar{\xi}_k  }_{~~~~~~~~~~~~\bar{\d'}_k}\right]\eta^{\a\b}}{(k^2+2 k\cdot p''-c'')^{13/2-j}}~;
	\end{split}
\end{equation}
\begin{equation}
	\begin{split}
		X'_{17} =-2 p^2 p_\d  \sum_{j=0}^2 \sum_{i=0}^{4-2j} M(j) \frac{\left[A^i B^{4-2j-i}\{p_i k_{7-2j-i}\zeta_{2j}\}^{\theta \d'  \o \d \bar{\chi}_k \bar{\xi}_k}_{~~~~~~~~~\bar{\d'_k}}\right]\eta^{\a\b}}{(k^2+2 k\cdot p''-c'')^{13/2-j}}~;
	\end{split}
\end{equation}
\\
\begin{equation}
	\begin{split}
		X'_{18} = p^2  \sum_{j=0}^1 \sum_{i=0}^{3-2j} M(j) \frac{\left[A^i B^{3-2j-i}\{p_i k_{8-2j-i}\zeta_{2j}\}^{\theta \o \d' \bar{\chi}_k \bar{\xi}_k}_{~~~~~~~~\bar{\d'_k}}\right]\eta^{\a\b}}{(k^2+2 k\cdot p''-c'')^{13/2-j}}~;
	\end{split}
\end{equation}
\begin{equation}
	\begin{split}
		X'_{19} =&-2 p^2 p_\d \sum_{j=0}^1 \sum_{i=0}^{3-2j} M(j) \frac{\left[A^i B^{3-2j-i}\{p_i k_{7-2j-i}\zeta_{2j}\}^{\theta \o \d' \bar{\chi_k} \bar{\xi}_k \bar{\d}_k}_{~~~~~~~~~~\bar{\d'_k}}\right]\eta^{\a\b}}{(k^2+2 k\cdot p''-c'')^{13/2-j}}~;
	\end{split}
\end{equation}
\begin{equation}
	\begin{split}
		X'_{1,10} =& p^2 p^2  \sum_{j=0}^1 \sum_{i=0}^{3-2j} M(j) \frac{\left[A^i B^{3-2j-i}\{p_i k_{6-2j-i}\zeta_{2j}\}^{\theta \o\d' \bar{\chi}_k \bar{\xi}_k}_{~~~~~~~~\bar{\d'_k}}\right]\eta^{\a\b}}{(k^2+2 k\cdot p''-c'')^{13/2-j}}~;
	\end{split}
\end{equation}

It is possible to repeat the procedure to determine the others $X'_{2l'}$:
\begin{equation}
	\begin{split}
		X'_{20} =-\eta_{\phi \phi'}  p^\xi  \sum_{j=0}^2 \sum_{i=0}^{4-2j}  \frac{\left[M(j) A^i B^{4-2j-i}\{p_i k_{9-2j-i}\zeta_{2j}\}^{\theta \o \phi \phi' \bar{\chi}_k}\right] \eta^{\a\b}}{(k^2+2 k\cdot p''-c'')^{13/2-j}};
		~.
	\end{split}
\end{equation}
\begin{equation}
	\begin{split}
		X'_{21}= &-\eta_{\phi \phi'}  p^\o  \sum_{j=0}^1 \sum_{i=0}^{3-2j} M(j) \frac{\left[A^i B^{3-2j-i}\{p_i k_{9-2j-i}\zeta_{2j}\}^{\theta \phi \phi'  \bar{\chi}_k \bar{\xi}_k}\right]\eta^{\a\b}}{(k^2+2 k\cdot p''-c'')^{13/2-j}};
	\end{split}
\end{equation}
\\
\begin{equation}
	\begin{split}
		X'_{22}= &\eta_{\phi \phi'} p^\o p^\xi \sum_{j=0}^1 \sum_{i=0}^{3-2j} M(j) \frac{\left[A^i B^{3-2j-i}\{p_i k_{8-2j-i}\zeta_{2j}\}^{\theta \phi \phi'  \bar{\chi}_k }\right]\eta^{\a\b}}{(k^2+2 k\cdot p''-c'')^{13/2-j}};
	\end{split}
\end{equation}
\\
\begin{equation}
	\begin{split}
		X'_{23}=&2\eta_{\phi \phi'}  p_\d  \sum_{j=0}^2 \sum_{i=0}^{4-2j} M(j) \frac{\left[A^i B^{4-2j-i}\{p_i k_{9-2j-i}\zeta_{2j}\}^{\theta \o \phi \phi'  \bar{\chi}_k\bar{\xi}_k \bar{\d}_k}\right]\eta^{\a\b}}{(k^2+2 k\cdot p''-c'')^{13/2-j}};
	\end{split}
\end{equation}
\\
\begin{equation}
	\begin{split}
		X'_{24} =& -\eta_{\phi \phi'} p^2  \sum_{j=0}^2 \sum_{i=0}^{4-2j} M(j)\frac{\left[A^i B^{4-2j-i}\{p_i k_{8-2j-i}\zeta_{2j}\}^{\theta \o \phi \phi' \bar{\xi}_k \bar{\chi}_k}\right]\eta^{\a\b}}{(k^2+2 k\cdot p''-c'')^{13/2-j}};
	\end{split}
\end{equation}
\\
\begin{equation}
	\begin{split}
		X'_{25} =& 2 p_\d  \sum_{j=0}^1 \sum_{i=0}^{3-2j} M(j) \frac{\left[A^i B^{3-2j-i}\{p_i k_{9-2j-i}\zeta_{2j}\}^{\theta   \o \d \bar{\chi}_k \bar{\xi}_k}\right]\eta^{\a\b}}{(k^2+2 k\cdot p''-c'')^{13/2-j}} ;
	\end{split}
\end{equation}
\\
\begin{equation}
	\begin{split}
		X'_{26} =& -4  p_\d p_{\d'} \sum_{j=0}^1 \sum_{i=0}^{3-2j} M(j) \frac{\left[A^i B^{3-2j-i}\{p_i k_{8-2j-i}\zeta_{2j}\}^{\theta \o \d  \bar{\chi}_k \bar{\xi}_k \bar{\d'}_k }\right]\eta^{\a\b}}{(k^2+2 k\cdot p''-c'')^{13/2-j}};
	\end{split}
\end{equation}
\\
\begin{equation}
	\begin{split}
		X'_{27} =&2 p^2 p_\d  \sum_{j=0}^1 \sum_{i=0}^{3-2j} M(j) \frac{\left[A^i B^{3-2j-i}\{p_i k_{7-2j-i}\zeta_{2j}\}^{\theta  \o \d \bar{\chi}_k \bar{\xi}_k}\right]\eta^{\a\b}}{(k^2+2 k\cdot p''-c'')^{13/2-j}};\\[3ex]
	\end{split}
\end{equation}
\begin{equation}
	\begin{split}
		X'_{28} =&- p^2  \sum_{j=0}^1 \sum_{i=0}^{2-2j} M(j) \frac{\left[A^i B^{2-2j-i}\{p_i k_{8-2j-i}\zeta_{2j}\}^{\theta \o  \bar{\chi}_k \bar{\xi}_k}\right]\eta^{\a\b}}{(k^2+2 k\cdot p''-c'')^{13/2-j}};
		\\[3ex]
	\end{split}
\end{equation}
\begin{equation}
	\begin{split}
		X'_{29} =&2 p^2 p_\d \sum_{j=0}^1 \sum_{i=0}^{2-2j} M(j) \frac{\left[A^i B^{2-2j-i}\{p_i k_{7-2j-i}\zeta_{2j}\}^{\theta \o  \bar{\chi}_k \bar{\xi}_k \bar{\d}_k}\right]\eta^{\a\b}}{(k^2+2 k\cdot p''-c'')^{13/2-j}};
	\end{split}
\end{equation}\\
\begin{equation}
	\begin{split}
		X'_{2,10} =&-p^2 p^2  \sum_{j=0}^1 \sum_{i=0}^{2-2j} M(j) \frac{\left[A^i B^{2-2j-i}\{p_i k_{6-2j-i}\zeta_{2j}\}^{\theta \o \bar{\chi}_k \bar{\xi}_k}\right]\eta^{\a\b}}{(k^2+2 k\cdot p''-c'')^{13/2-j}};
	\end{split}
\end{equation}

The structure of $X'_{3l'}$ is the same as that of $X'_{0l'}$, except for a $-1$ factor and for the tensor indices $\alpha$ and $\beta$, which are in a $J_r$ function instead of a metric tensor. This also occurs with $X'_{4l'}, X'_{5l'}$, and $X'_{1l'}$, as well as with $X'_{6l'}$ and $X'_{2l'}$. Therefore, we are now presenting the remaining $X'_{ll'}$:
\\[5ex]
$X'_{3l'}:$
\begin{equation}
	\begin{split}
		X'_{30} =\eta_{\phi \phi'}  p^\xi  \sum_{j=0}^3 \sum_{i=0}^{6-2j}  \frac{\left[M(j) A^i B^{6-2j-i}\{p_i k_{9-2j-i}\zeta_{2j}\}^{\theta \o \phi \phi' \a\b \bar{\chi}_k}\right] }{(k^2+2 k\cdot p''-c'')^{13/2-j}}
		~.
	\end{split}
\end{equation}
\begin{equation}
	\begin{split}
		X'_{31}= &\eta_{\phi \phi'} p^\o  \sum_{j=0}^2 \sum_{i=0}^{5-2j} M(j) \frac{\left[A^i B^{5-2j-i}\{p_i k_{9-2j-i}\zeta_{2j}\}^{\theta \phi \phi' \a\b \bar{\chi}_k \bar{\xi}_k}\right]}{(k^2+2 k\cdot p''-c'')^{13/2-j}}
	\end{split}
\end{equation}
\begin{equation}
	\begin{split}
		X'_{32}= -\eta_{\phi \phi'} p^\o p^\xi \sum_{j=0}^2 \sum_{i=0}^{5-2j} M(j) \frac{\left[A^i B^{5-2j-i}\{p_i k_{8-2j-i}\zeta_{2j}\}^{\theta \phi \phi' \a\b \bar{\chi}_k }\right]}{(k^2+2 k\cdot p''-c'')^{13/2-j}};
	\end{split}
\end{equation}
\begin{equation}
	\begin{split}
		X'_{33}=&-2\eta_{\phi \phi'} p_\d  \sum_{j=0}^3 \sum_{i=0}^{6-2j} M(j) \frac{\left[A^i B^{6-2j-i}\{p_i k_{9-2j-i}\zeta_{2j}\}^{\theta \o \phi \phi' \a\b \bar{\chi}_k\bar{\xi}_k \bar{\d}_k}\right]}{(k^2+2 k\cdot p''-c'')^{13/2-j}};
	\end{split}
\end{equation}
\begin{equation}
	\begin{split}
		X'_{34} =\eta_{\phi \phi'} p^2  \sum_{j=0}^3 \sum_{i=0}^{6-2j} M(j)\frac{\left[A^i B^{6-2j-i}\{p_i k_{8-2j-i}\zeta_{2j}\}^{\theta \o \phi \phi' \a\b \bar{\xi}_k \bar{\chi}_k}\right]}{(k^2+2 k\cdot p''-c'')^{13/2-j}};
	\end{split}
\end{equation}
\begin{equation}
	\begin{split}
		X'_{35} =&- 2 p_\d  \sum_{j=0}^2 \sum_{i=0}^{5-2j} M(j) \frac{\left[A^i B^{5-2j-i}\{p_i k_{9-2j-i}\zeta_{2j}\}^{\theta \a\b  \o \d \bar{\chi}_k \bar{\xi}_k}\right]}{(k^2+2 k\cdot p''-c'')^{13/2-j}}; \\[2ex]
	\end{split}
\end{equation}
\begin{equation}
	\begin{split}
		X'_{36} =& 4   p_\d p_{\d'} \sum_{j=0}^2 \sum_{i=0}^{5-2j} M(j) \frac{\left[A^i B^{5-2j-i}\{p_i k_{8-2j-i}\zeta_{2j}\}^{\theta \o \a\b \d  \bar{\chi}_k \bar{\xi}_k \bar{\d'}_k }\right]}{(k^2+2 k\cdot p''-c'')^{13/2-j}};
	\end{split}
\end{equation}
\begin{equation}
	\begin{split}
		X'_{37} =&-2 p^2 p_\d  \sum_{j=0}^2 \sum_{i=0}^{5-2j} M(j) \frac{\left[A^i B^{5-2j-i}\{p_i k_{7-2j-i}\zeta_{2j}\}^{\theta \a\b \o \d \bar{\chi}_k \bar{\xi}_k}\right]}{(k^2+2 k\cdot p''-c'')^{13/2-j}};\\[3ex]
	\end{split}
\end{equation}
\begin{equation}
	\begin{split}
		X'_{38} =& p^2  \sum_{j=0}^2 \sum_{i=0}^{4-2j} M(j) \frac{\left[A^i B^{4-2j-i}\{p_i k_{8-2j-i}\zeta_{2j}\}^{\theta \o \a\b \bar{\chi}_k \bar{\xi}_k}\right]}{(k^2+2 k\cdot p''-c'')^{13/2-j}};
		\\[3ex]
	\end{split}
\end{equation}
\begin{equation}
	\begin{split}
		X'_{39} =&-2p^2 p_\d \sum_{j=0}^2 \sum_{i=0}^{4-2j} M(j) \frac{\left[A^i B^{4-2j-i}\{p_i k_{7-2j-i}\zeta_{2j}\}^{\theta \o \a\b \bar{\chi}_k \bar{\xi}_k}\bar{\d}_k\right]}{(k^2+2 k\cdot p''-c'')^{13/2-j}};
	\end{split}
\end{equation}
\begin{equation}
	\begin{split}
		X'_{3,10} =&p^2 p^2  \sum_{j=0}^1 \sum_{i=0}^{2-2j} M(j) \frac{\left[A^i B^{2-2j-i}\{p_i k_{6-2j-i}\zeta_{2j}\}^{\theta \o \a\b  \bar{\chi}_k \bar{\xi}_k}\right]}{(k^2+2 k\cdot p''-c'')^{13/2-j}};
	\end{split}
\end{equation}
\\[6ex]
$X'_{4l'}:$
\begin{equation}
	\begin{split}
		X'_{40} = &-\eta_{\phi \phi'}  p^\xi  \sum_{j=0}^2 \sum_{i=0}^{5-2j} M(j) \frac{\left[A^i B^{5-2j-i}\{p_i k_{9-2j-i}\zeta_{2j}\}^{\theta \o \phi \phi' \a \bar{\chi_k}\bar{\b}_k}\right]}{(k^2+2 k\cdot p''-c'')^{13/2-j}}~;
	\end{split}
\end{equation}
\begin{equation}
	\begin{split}
		X'_{41}= &-\eta_{\phi \phi'}  p^\o  \sum_{j=0}^2 \sum_{i=0}^{4-2j} M(j)\frac{\left[A^i B^{4-2j-i}\{p_i k_{9-2j-i}\zeta_{2j}\}^{\theta \phi \phi' \a \bar{\chi}_k \bar{\xi}_k \bar{\b_k}}\right]}{(k^2+2 k\cdot p''-c'')^{13/2-j}}~~;
	\end{split}
\end{equation}
\begin{equation}
	\begin{split}
		X'_{42}= &\eta_{\phi \phi'}  p^\o p^\xi \sum_{j=0}^2 \sum_{i=0}^{4-2j} M(j) \frac{\left[A^i B^{4-2j-i}\{p_i k_{8-2j-i}\zeta_{2j}\}^{\theta \phi \phi' \a \bar{\chi}_k \bar{\b_k} }\right]}{(k^2+2 k\cdot p''-c'')^{13/2-j}}~;
	\end{split}
\end{equation}
\\
\begin{equation}
	\begin{split}
		X'_{43}= &2\eta_{\phi \phi'}  p_{\d} \sum_{j=0}^2 \sum_{i=0}^{5-2j} M(j) \frac{\left[A^i B^{5-2j-i}\{p_i k_{8-2j-i}\zeta_{2j}\}^{\theta \o \phi \phi' \a \bar{\chi}_k \bar{\xi}_k \bar{\d}_k \bar{\b}_k}\right]}{(k^2+2 k\cdot p''-c'')^{13/2-j}}~~; 
	\end{split}
\end{equation}
\\
\begin{equation}
	\begin{split}
		X'_{44}= &-\eta_{\phi \phi'}  p^2 \sum_{j=0}^2 \sum_{i=0}^{5-2j} M(j) \frac{\left[A^i B^{5-2j-i}\{p_i k_{8-2j-i}\zeta_{2j}\}^{\theta \o \phi \phi' \a \bar{\chi}_k \bar{\xi}_k  \bar{\b}_k}\right]}{(k^2+2 k\cdot p''-c'')^{13/2-j}}~~; 
	\end{split}
\end{equation}
\\
\begin{equation}
	\begin{split}
		X'_{45}= & 2 p_\d  \sum_{j=0}^2 \sum_{i=0}^{4-2j} M(j) \frac{\left[A^i B^{4-2j-i}\{p_i k_{9-2j-i}\zeta_{2j}\}^{\theta \o \d \a \bar{\chi}_k \bar{\xi}_k \bar{\b}_k}\right]}{(k^2+2 k\cdot p''-c'')^{13/2-j}}~; 
	\end{split}
\end{equation}
\\
\begin{equation}
	\begin{split}
		X'_{46} =& -4   p_\d p_{\d''} \sum_{j=0}^2 \sum_{i=0}^{4-2j} M(j) \frac{\left[A^i B^{4-2j-i}\{p_i k_{8-2j-i}\zeta_{2j}\}^{\theta \o  \d \a \bar{\d''_k} \bar{\chi}_k \bar{\xi}_k \bar{\b}_k }\right]}{(k^2+2 k\cdot p''-c'')^{13/2-j}}~;
	\end{split}
\end{equation}
\begin{equation}
	\begin{split}
		X'_{47} =2 p^2 p_\d  \sum_{j=0}^2 \sum_{i=0}^{4-2j} M(j) \frac{\left[A^i B^{4-2j-i}\{p_i k_{7-2j-i}\zeta_{2j}\}^{\theta \a  \o \d \bar{\chi}_k \bar{\xi}_k \bar{\b}_k}\right]}{(k^2+2 k\cdot p''-c'')^{13/2-j}}~;
	\end{split}
\end{equation}
\\
\begin{equation}
	\begin{split}
		X'_{48} = -p^2  \sum_{j=0}^1 \sum_{i=0}^{3-2j} M(j) \frac{\left[A^i B^{3-2j-i}\{p_i k_{8-2j-i}\zeta_{2j}\}^{\theta \o \a \bar{\chi}_k \bar{\xi}_k\bar{\b}_k}\right]}{(k^2+2 k\cdot p''-c'')^{13/2-j}}~;
	\end{split}
\end{equation}
\begin{equation}
	\begin{split}
		X'_{49} =&2 p^2 p_\d \sum_{j=0}^1 \sum_{i=0}^{3-2j} M(j) \frac{\left[A^i B^{3-2j-i}\{p_i k_{7-2j-i}\zeta_{2j}\}^{\theta \o \a \bar{\chi_k} \bar{\xi}_k \bar{\d}_k \bar{\b}_k}\right]}{(k^2+2 k\cdot p''-c'')^{13/2-j}}~;
	\end{split}
\end{equation}
\begin{equation}
	\begin{split}
		X'_{4,10} =& -p^2 p^2  \sum_{j=0}^1 \sum_{i=0}^{3-2j} M(j) \frac{\left[A^i B^{3-2j-i}\{p_i k_{6-2j-i}\zeta_{2j}\}^{\theta \o\a \bar{\chi}_k \bar{\xi}_k \bar{\b}_k}\right]}{(k^2+2 k\cdot p''-c'')^{13/2-j}}~;
	\end{split}
\end{equation}
\\[6ex]
$X'_{5l'}:$
\begin{equation}
	\begin{split}
		X'_{50} = &-\eta_{\phi \phi'}  p^\xi  \sum_{j=0}^2 \sum_{i=0}^{5-2j} M(j) \frac{\left[A^i B^{5-2j-i}\{p_i k_{9-2j-i}\zeta_{2j}\}^{\theta \o \phi \phi' \b \bar{\chi_k}\bar{\a}_k}\right]}{(k^2+2 k\cdot p''-c'')^{13/2-j}}~;
	\end{split}
\end{equation}
\begin{equation}
	\begin{split}
		X'_{51}= &-\eta_{\phi \phi'}  p^\o  \sum_{j=0}^2 \sum_{i=0}^{4-2j} M(j)\frac{\left[A^i B^{4-2j-i}\{p_i k_{9-2j-i}\zeta_{2j}\}^{\theta \phi \phi' \b \bar{\chi}_k \bar{\xi}_k \bar{\a_k}}\right]}{(k^2+2 k\cdot p''-c'')^{13/2-j}}~~;
	\end{split}
\end{equation}
\begin{equation}
	\begin{split}
		X'_{52}= &\eta_{\phi \phi'}  p^\o p^\xi \sum_{j=0}^2 \sum_{i=0}^{4-2j} M(j) \frac{\left[A^i B^{4-2j-i}\{p_i k_{8-2j-i}\zeta_{2j}\}^{\theta \phi \phi' \b \bar{\chi}_k \bar{\a_k} }\right]}{(k^2+2 k\cdot p''-c'')^{13/2-j}}~;
	\end{split}
\end{equation}
\\
\begin{equation}
	\begin{split}
		X'_{53}= &2\eta_{\phi \phi'}  p_{\d} \sum_{j=0}^2 \sum_{i=0}^{5-2j} M(j) \frac{\left[A^i B^{5-2j-i}\{p_i k_{8-2j-i}\zeta_{2j}\}^{\theta \o \phi \phi' \b \bar{\chi}_k \bar{\xi}_k \bar{\d}_k \bar{\a}_k}\right]}{(k^2+2 k\cdot p''-c'')^{13/2-j}}~~; 
	\end{split}
\end{equation}
\\
\begin{equation}
	\begin{split}
		X'_{54}= &-\eta_{\phi \phi'}  p^2 \sum_{j=0}^2 \sum_{i=0}^{5-2j} M(j) \frac{\left[A^i B^{5-2j-i}\{p_i k_{8-2j-i}\zeta_{2j}\}^{\theta \o \phi \phi' \b \bar{\chi}_k \bar{\xi}_k  \bar{\a}_k}\right]}{(k^2+2 k\cdot p''-c'')^{13/2-j}}~~; 
	\end{split}
\end{equation}
\\
\begin{equation}
	\begin{split}
		X'_{55}= & 2 p_\d  \sum_{j=0}^2 \sum_{i=0}^{4-2j} M(j) \frac{\left[A^i B^{4-2j-i}\{p_i k_{9-2j-i}\zeta_{2j}\}^{\theta \o \d \b \bar{\chi}_k \bar{\xi}_k \bar{\a}_k}\right]}{(k^2+2 k\cdot p''-c'')^{13/2-j}}~; 
	\end{split}
\end{equation}
\\
\begin{equation}
	\begin{split}
		X'_{56} =& -4   p_\d p_{\d''} \sum_{j=0}^2 \sum_{i=0}^{4-2j} M(j) \frac{\left[A^i B^{4-2j-i}\{p_i k_{8-2j-i}\zeta_{2j}\}^{\theta \o  \d \b \bar{\d''_k} \bar{\chi}_k \bar{\xi}_k \bar{\a}_k }\right]}{(k^2+2 k\cdot p''-c'')^{13/2-j}}~;
	\end{split}
\end{equation}
\begin{equation}
	\begin{split}
		X'_{57} =2 p^2 p_\d  \sum_{j=0}^2 \sum_{i=0}^{4-2j} M(j) \frac{\left[A^i B^{4-2j-i}\{p_i k_{7-2j-i}\zeta_{2j}\}^{\theta \b  \o \d \bar{\chi}_k \bar{\xi}_k \bar{\a}_k}\right]}{(k^2+2 k\cdot p''-c'')^{13/2-j}}~;
	\end{split}
\end{equation}
\\
\begin{equation}
	\begin{split}
		X'_{58} = -p^2  \sum_{j=0}^1 \sum_{i=0}^{3-2j} M(j) \frac{\left[A^i B^{3-2j-i}\{p_i k_{8-2j-i}\zeta_{2j}\}^{\theta \o \b \bar{\chi}_k \bar{\xi}_k\bar{\a}_k}\right]}{(k^2+2 k\cdot p''-c'')^{13/2-j}}~;
	\end{split}
\end{equation}
\begin{equation}
	\begin{split}
		X'_{59} =&2 p^2 p_\d \sum_{j=0}^1 \sum_{i=0}^{3-2j} M(j) \frac{\left[A^i B^{3-2j-i}\{p_i k_{7-2j-i}\zeta_{2j}\}^{\theta \o \b \bar{\chi_k} \bar{\xi}_k \bar{\d}_k \bar{\a}_k}\right]}{(k^2+2 k\cdot p''-c'')^{13/2-j}}~;
	\end{split}
\end{equation}
\begin{equation}
	\begin{split}
		X'_{5,10} =& -p^2 p^2  \sum_{j=0}^1 \sum_{i=0}^{3-2j} M(j) \frac{\left[A^i B^{3-2j-i}\{p_i k_{6-2j-i}\zeta_{2j}\}^{\theta \o\b \bar{\chi}_k \bar{\xi}_k \bar{\a}_k}\right]}{(k^2+2 k\cdot p''-c'')^{13/2-j}}~;
	\end{split}
\end{equation}
\\[6ex]
$X'_{6l'}:$
\begin{equation}
	\begin{split}
		X'_{60} = p^\xi  \sum_{j=0}^2 \sum_{i=0}^{4-2j}  \frac{\left[M(j) A^i B^{4-2j-i}\{p_i k_{9-2j-i}\zeta_{2j}\}^{\theta \o  \bar{\chi}_k \bar{\a}_k \bar{\b}_k}\right] }{(k^2+2 k\cdot p''-c'')^{13/2-j}};
		~.
	\end{split}
\end{equation}
\begin{equation}
	\begin{split}
		X'_{61}= &  p^\o  \sum_{j=0}^1 \sum_{i=0}^{3-2j} M(j) \frac{\left[A^i B^{3-2j-i}\{p_i k_{9-2j-i}\zeta_{2j}\}^{\theta  \bar{\chi}_k \bar{\xi}_k \bar{\a}_k \bar{\b}_k}\right]}{(k^2+2 k\cdot p''-c'')^{13/2-j}};
	\end{split}
\end{equation}
\\
\begin{equation}
	\begin{split}
		X'_{62}= &- p^\o p^\xi \sum_{j=0}^1 \sum_{i=0}^{3-2j} M(j) \frac{\left[A^i B^{3-2j-i}\{p_i k_{8-2j-i}\zeta_{2j}\}^{\theta   \bar{\chi}_k \bar{\a}_k \bar{\b}_k}\right]}{(k^2+2 k\cdot p''-c'')^{13/2-j}};
	\end{split}
\end{equation}
\\
\begin{equation}
	\begin{split}
		X'_{63}=&-2 p_\d  \sum_{j=0}^2 \sum_{i=0}^{4-2j} M(j) \frac{\left[A^i B^{4-2j-i}\{p_i k_{9-2j-i}\zeta_{2j}\}^{\theta \o  \bar{\chi}_k\bar{\xi}_k \bar{\d}_k \bar{\a}_k \bar{\b}_k}\right]}{(k^2+2 k\cdot p''-c'')^{13/2-j}};
	\end{split}
\end{equation}
\\
\begin{equation}
	\begin{split}
		X'_{64} =&  p^2  \sum_{j=0}^2 \sum_{i=0}^{4-2j} M(j)\frac{\left[A^i B^{4-2j-i}\{p_i k_{8-2j-i}\zeta_{2j}\}^{\theta \o  \bar{\xi}_k \bar{\chi}_k \bar{\a}_k \bar{\b}_k}\right]}{(k^2+2 k\cdot p''-c'')^{13/2-j}};
	\end{split}
\end{equation}
\\
\begin{equation}
	\begin{split}
		X'_{65} =& -2 p_\d  \sum_{j=0}^1 \sum_{i=0}^{3-2j} M(j) \frac{\left[A^i B^{3-2j-i}\{p_i k_{9-2j-i}\zeta_{2j}\}^{\theta  \o \d \bar{\chi}_k \bar{\xi}_k \bar{\a}_k \bar{\b}_k}\right]}{(k^2+2 k\cdot p''-c'')^{13/2-j}} ;
	\end{split}
\end{equation}
\\
\begin{equation}
	\begin{split}
		X'_{66} =& 4  p_\d p_{\d'} \sum_{j=0}^1 \sum_{i=0}^{3-2j} M(j) \frac{\left[A^i B^{3-2j-i}\{p_i k_{8-2j-i}\zeta_{2j}\}^{\theta \o \d  \bar{\chi}_k \bar{\xi}_k \bar{\d'}_k \bar{\a}_k \bar{\b}_k}\right]}{(k^2+2 k\cdot p''-c'')^{13/2-j}};
	\end{split}
\end{equation}
\\
\begin{equation}
	\begin{split}
		X'_{67} =&-2 p^2 p_\d  \sum_{j=0}^1 \sum_{i=0}^{3-2j} M(j) \frac{\left[A^i B^{3-2j-i}\{p_i k_{7-2j-i}\zeta_{2j}\}^{\theta  \o \d \bar{\chi}_k \bar{\xi}_k \bar{\a}_k \bar{\b}_k}\right]}{(k^2+2 k\cdot p''-c'')^{13/2-j}};\\[3ex]
	\end{split}
\end{equation}
\begin{equation}
	\begin{split}
		X'_{68} =& p^2  \sum_{j=0}^1 \sum_{i=0}^{2-2j} M(j) \frac{\left[A^i B^{2-2j-i}\{p_i k_{8-2j-i}\zeta_{2j}\}^{\theta \o  \bar{\chi}_k \bar{\xi}_k \bar{\a}_k \bar{\b}_k}\right]}{(k^2+2 k\cdot p''-c'')^{13/2-j}};
		\\[3ex]
	\end{split}
\end{equation}
\begin{equation}
	\begin{split}
		X'_{69} =&-2 p^2 p_\d \sum_{j=0}^1 \sum_{i=0}^{2-2j} M(j) \frac{\left[A^i B^{2-2j-i}\{p_i k_{7-2j-i}\zeta_{2j}\}^{\theta \o  \bar{\chi}_k \bar{\xi}_k  \bar{\d}_k\bar{\a}_k  \bar{\b}_k}\right]}{(k^2+2 k\cdot p''-c'')^{13/2-j}};
	\end{split}
\end{equation}\\
\begin{equation}
	\begin{split}
		X'_{6,10} =&p^2 p^2  \sum_{j=0}^1 \sum_{i=0}^{2-2j} M(j) \frac{\left[A^i B^{2-2j-i}\{p_i k_{6-2j-i}\zeta_{2j}\}^{\theta \o \bar{\chi}_k \bar{\xi}_k \bar{\a}_k  \bar{\b}_k}\right]}{(k^2+2 k\cdot p''-c'')^{13/2-j}}.
	\end{split}
\end{equation}

\bibliographystyle{ieeetr}
\bibliography{refs}

\end{document}